\documentstyle[preprint,aps,eqsecnum]{revtex}
\tighten
\begin{document}
\draft
\title{{\bf  NEW ASPECTS OF REHEATING}\footnote{To appear in the Proceedings of
the School on String Gravity and
Physics at the Planck Scale, Erice, Sicily 8 September 1995, Editor: N.
Sanchez}}
\author{{\bf D. Boyanovsky$^{(a)}$,
M. D'Attanasio\footnote{Della Riccia fellow}$^{(b)(d)}$,
H.J. de Vega$^{(b)}$,
\\ R. Holman$^{(c)}$ and
 D.-S. Lee$^{(e)}$}}
\address
{ (a)  Department of Physics and Astronomy, University of
Pittsburgh, Pittsburgh, PA. 15260, U.S.A. \\
 (b)  Laboratoire de Physique Th\'eorique et Hautes
Energies$^{[\dag]}$
Universit\'e Pierre et Marie Curie (Paris VI) ,
Tour 16, 1er. \'etage, 4, Place Jussieu
75252 Paris cedex 05, France \\
 (c) Department of Physics, Carnegie Mellon University,
Pittsburgh,
PA. 15213, U. S. A. \\
 (d) I.N.F.N., Gruppo Collegato di Parma, Italy\\
(e) Department of Physics and Astronomy, University of North Carolina\\
Chapel Hill, N.C. 27599, U.S.A.}
\date{ }
\maketitle
\begin{abstract}

The reheating stage in post-inflationary cosmologies is reanalyzed.
New techniques from
non-equilibrium quantum field theory allow a consistent derivation of
the equation of motion
including the non-linearity of the dynamics. These offer a rationale for the
elementary  theory of  reheating based on
single particle decay which is seen to be valid only in the linear regime
of coherent oscillations of the scalar field.

A new non-perturbative mechanism of induced
amplification of quantum fluctuations is introduced and studied in detail,
both analytically and
numerically. This is a non-linear mechanism that is typically a  far more
efficient way of transfering energy out of the zero mode and into production
of lighter particles than single particle decay. Thermalization is discussed
and we estimate  the reheating temperature to be
of the order of the inflaton mass, thus providing a potential solution to the
Polonyi and moduli problems.

\end{abstract}

\section{\bf Introduction}

In typical inflationary scenarios of the early universe, after
the many e-folds of inflation necessary to solve the horizon and homogeneity
problems, the matter and radiation energy density had been red-shifted to
almost zero. However, once inflation has ended, this scenario has to merge with
the standard big bang, radiation dominated cosmology. The succesful
intertwining of inflation with the standard big bang cosmology thus
necessitates some source of energy and entropy to rethermalize, or reheat the
universe\cite{revs,linde,rocky}.

Models that reheat to temperatures above the GUT scale would have to cope
with the
monopole and domain wall and inhomogeneity problems associated with the
breakdown of
their symmetries. However there is a fundamental
 scale for a reheating temperature: nucleosynthesis,
this requires reheating temperatures in excess of 10-100 Mev. If baryogenesis
can occur at
the electroweak scale then it would be imperative to have reheating
temperatures larger than
say a few hundred Gev (a baryon over antibaryon excess may be wiped out if
this occurs
during the fast inflationary era). Therefore, nucleosynthesis and particle
physics restrict the
reheating temperature but leave a wide window open for the acceptable values
of the
reheating temperatures: certainly larger than about 100 Mev,  and desirably,
larger than say about 1 Tev but not close to GUT scale.

The standard picture\cite{linde,rocky,dolgov1} of old and new inflation
invokes
a scalar field, the inflaton, that produces a phase transition (either second
or first order) and that at the end of inflation the expectation value of its
zero momentum mode oscillates about the minimum of its (effective)
potential. In chaotic inflation\cite{chaotic}, the inflaton energy density
still drives an inflationary phase, but there  need not be any phase
transition for this to occur. This inflaton field is
constrained\cite{revs,linde,rocky} to have a mass a few orders of magnitude
smaller than the Planck mass, but typically much larger than the masses of the
particles involved in the particle physics models. In the standard view of
inflationary models, the expectation value of the inflaton field oscillates
around the minimum of its potential, and its couplings to the lighter particles
allow the inflaton field to decay. This decay process is then supposed to
induce a damping term in the evolution equation for the inflaton expectation
value of the form $\Gamma \dot{\phi}$ with $\Gamma$ being the total decay rate
of the inflaton field\cite{linde,abbott,kirilova}. The standard estimate for
the reheating temperature based on single-particle decay\cite{linde} is then
obtained by comparing the total decay rate of the particle ($\Gamma$) to that
of the expansion, obtaining $T_r \simeq 0.1\sqrt{\Gamma M_{pl}}$\cite{linde}.

Recent investigations of the non-linear quantum dynamics of scalar
fields reveal a variety of new and striking
phenomena\cite{tadpole,frwI,paris,kofman,branden3,reheating,erice1}.
The main relevant implication
for the reheating problem being that the particle production induced
by the time evolution of the inflaton is significatively {\bf
different} from linear estimates.

The non-linear (quantum) effects lead to an
extremely effective  dissipational dynamics and particle
production even in the simplest self-interacting scalar field
theory\cite{kofman,reheating} in which single particle decay is kinematically
forbidden. In the case in which only the {\it classical} evolution of the
expectation value is taken into account in the evolution of the quantum
fluctuations, this mechanism corresponds to parametric amplification or
parametric resonance, because the classical time evolution is periodic in
time \cite{dau}. However when back-reaction effects are taken into
account, particle
production damps out the evolution of the scalar field, and this damped
evolution is incorporated in the equations for the quantum fluctuations. In
this case the time evolution of the scalar field {\em including} the
dissipative effects is not periodic; this implies that the situation is far
different from parametric amplification. We refer to this situation, which
includes the back-reaction effects as {\bf induced amplification}, to
distinguish it from parametric resonant amplification \cite{dau}.

Back-reaction effects\cite{kofman,reheating} drastically change the picture of
particle production. In the case of parametric amplification, particle
production never shuts off and the total number of particles created is
infinite in Minkowski. It is only when the back-reaction is incorporated
that the damping
effects on the expectation value feed back to the particle production
mechanism, eventually shutting it off. This has been studied in detail
analytically and numerically in\cite{reheating} for a self-interacting scalar
field.

Induced amplification is most effective when the amplitude of the expectation
value of the scalar field (from the equilibrium value) is large and is,
therefore, an intrinsically non-linear process, which we refer to as {\it
non-linear relaxation}.

Particle production via parametric amplification of quantum fluctuations has
been studied recently in the semiclassical approximation, but without
back-reaction effects, in connection with fermion production by (pseudo)
Nambu-Goldstone bosons\cite{freese}.

Induced amplification and non-linear relaxation will be the primary mechanism
for dissipation via particle production in the case of large amplitude of the
scalar field. Such is the case in chaotic infation scenarios\cite{chaotic}, as
well as in the out of equilibrium regime during phase transitions for fields
with very flat potentials near the maximum, which is typically required for a
long inflationary phase and also occurs for the moduli fields in string
theory\cite{moduli}. In these inflationary models with light moduli fields,
with masses
typically  of the order of a Tev, and couplings of order ($Tev/M_{Pl}$) the
standard reheating
theory predicts reheating temperatures of the order of 0.01Mev, well below
nucleosynthesis
temperatures, this is the Polonyi problem\cite{moduli,banks2,lyth,randall}.

The  non-linear mechanism of induced amplification of fluctuations
and copious particle
production in the large amplitude regime of the scalar field
will necessarily modify the standard picture of reheating.
The standard reheating scenario
was primarily
based on the premise that particle production only ocurred when the inflaton
oscillates with small amplitude at the bottom of the (effective) potential.
Understanding the time scales and non-linear processes of reheating and
eventually thermalization acquires further importance with the (speculative)
possibility that
asymptotic oscillations of the inflaton field around the minima of the
(effective) potential may still be present in today's universe in the form of
dark matter\cite{kofman,hogan}.

Thermalization is a process that is fundamentally different from particle
production. Typically, as will be seen below, the particles produced by the
process of induced amplification will be in a non-thermal
distribution\cite{reheating}. Thermalization is a collisional process, in
which the produced particles will exchange energy (and momentum) and eventually
achieve a thermal distribution. In principle the time scales for dissipation
via particle production may be widely different from the time scales of
collisional thermalization. For very weakly coupled theories, if the
distribution of produced particles via induced amplification is very far from
thermal, many collisions will be necessary to thermalize the system and the
time scale for thermalization may very well be large compared to the time scale
of dissipation via particle production. This will have implications in
cosmology discussed in the conclusions.

The process of thermalization, reheating and relaxation of perturbations, and
the ensuing production of entropy is also very relevant in heavy ion
collisions\cite{mueller} and in phase transitions in particle physics, at the
electroweak scale within the context of baryogenesis\cite{kaplan}, as well as
for the quark-gluon plasma, deconfining and chiral phase
transitions\cite{mueller,quiral}.

In the linear relaxation case the important scale is determined by the decay
rate $\Gamma$, usually referred to as the damping rate.
The literature\cite{landsman} usually identifies $\Gamma \approx
{\mbox{Im}\,}\Sigma(\omega,\vec{k})/ \omega$ with the
``thermalization rate''. Here
${\mbox{Im}\,}\Sigma(\omega,\vec{k})$ is the imaginary part of
the self-energy. In this
article we offer a {\em real time} critique of this relation by studying
the real time evolution and relaxation of linearized perturbations, and point
out the following observations:

\begin{enumerate}

\item{The damping rate is identical to the imaginary part of the self energy
{\it only} when the inflaton (the particle interpolated by the order parameter)
 is a resonance with $\Gamma$ being its width.
In the complex frequency plane this
corresponds to a pole in the second (unphysical) Riemann sheet. Even in this
case, this damping rate corresponds to the relaxation of the {\em expectation
value of the scalar field} and is exponential for some time regime. However,
eventually relaxation continues
with a power law tail. When the imaginary part is
zero on-shell and the one
particle pole is below the multiparticle
thresholds, relaxation is given by a power law and no damping rate can be
associated with the imaginary part of the self energy.
Moreover, the relation
$\Gamma \approx \mbox{Im}\,\Sigma(\omega,\vec{k})/\omega$
only holds in the {\bf linear} approximation around equilibrium.}

\item{Even when relaxation of the expectation value of the scalar field is
exponential, this damping rate determines the approach to equilibrium of the
{\it expectation value} of the scalar field but it cannot be
immediately inferred
that the same time scale describes thermalization, i.e. the approach to a
Bose Einstein distribution of an initial off-equilibrium distribution.
Thermalization is a rather different process and has to be described with a
Boltzmann equation with a collision term.  Thus, rather than interpreting this
time scale as a thermalization scale, we interpret it as a relaxation scale for
the expectation value.}

\item{We consider the non-linear quantum field evolution (non-linear
relaxation) incorporating the order parameter into the effective mass
in a self-consistent way. [The self-consistency requirement makes the evolution
equations non-linear].   We study the inflaton evolution in real time
coupled both to light  scalars and fermions, providing exhaustive
numerical results.  We find that the
non-linear relaxation time scales can be {\bf much shorter} than those
predicted by linear relaxation. In addition,  the particles are
produced with a
momentum distribution that is very far from thermal and skewed towards low
momentum. In the case of fermions we provide numerical evidence for the
phenomenon of Pauli blocking that hinders dissipation and production of
fermion-antifermion pairs.
This phenomenon is very similar to that found by Kluger and
collaborators\cite{kluger} in their studies of fermionic back-reaction in the
presence of strong electric fields. }

\item{We find that the reheating temperature is in between
the inflaton mass and the lighter scalar mass.
The precise relation is argued to depend on the product of the
coupling and the
initial amplitude of the inflaton zero mode and also logarithmically on
the coupling. The
distribution of produced particles is very far from thermal and skewed
towards momenta
of the order of or smaller than the inflaton mass. We argue that this
implies fairly long
thermalization times and a wide separation of the relevant time scales.}
\end{enumerate}

Both time scales will have to be understood in detail to provide a {\it
quantitative} and reliable estimate of the reheating temperature.

Although we are ultimately interested in describing reheating and
thermalization in Friedmann-Robertson-Walker
cosmologies, in this article we will work
in Minkowski space {\it assuming} that the relaxation time scales are much
shorter than the expansion time scale.  Furthermore, we want to isolate
dissipative effects arising from particle production from those resulting from
the red-shift in an expanding cosmology.

A more detailed  account of the results presented here
can be found in ref.\cite{prdpaper}.

In section II, we introduce the general model that we propose to study and
summarize the necessary ingredients of non-equilibrium field theory to provide
the reader with the essentials needed to reproduce our calculations,
as these do
not seem to be part of the standard techniques.
In section III, we study
linear relaxation in real time and elucidate the r\^ole of the imaginary parts
of self energies, their interpretation in real time and criticize the
identification with ``thermalization rates''. Here we provide a derivation
of the reheating
temperature based on single particle decay, valid in the linear regime of
coherent oscillations
of the scalar (inflaton) field. We also discuss the basics of the Polonyi
problem.

 In section IV we study the
process of non-linear relaxation, and particle production both for scalars and
fermions, providing exhaustive numerical results. Here we also discuss the
issue of thermalization, and the important time scales associated with this
collisional process. Under
the assumption of thermalization rates larger than the expansion rate we
provide an estimate
of the reheating temperature based on the total number of produced particles.

Finally we conclude in section V with a
discussion of the implications of our results, in particular we
 emphasize the possibility that
the new mechanism for reheating may provide a solution to the Polonyi problem.
 An Appendix is
devoted to a pedagogical exercise in non-equilibrium field theory.

\section{The Model and the Methods}

We consider the simplest model \cite{linde,rocky} where the inflaton field
$\Phi$ couples to a scalar $\sigma$ and to a fermion field $\psi$. That is,

\begin{equation}
{\cal{L} }= -{1 \over 2} \Phi \,( \partial^2 +m^2_{\Phi}+g \sigma^2)\,
\Phi -{ \lambda_{\Phi} \over 4!}  \, \Phi^4 -{1 \over 2} \sigma \,
(\partial^2 + m_{\sigma}^2)  \, \sigma
     -{ \lambda_{\sigma} \over 4!}  \, \sigma^4+
{\bar{\psi}} (i {\not\!{\partial}} -m_{\psi}
 -y \Phi  ) \psi \;. \nonumber 
\end{equation}
The case $m_{\sigma},m_{\psi} \ll m_{\Phi}$ will be of particular relevance,
since in this case there are open decay channels for the inflaton.

We will investigate the scalar and fermionic couplings independently.  Although
the situation for reheating corresponds to (almost) zero temperature, we will
study the case of linear relaxation at finite temperature. The reason for this
is that finite temperature effects reflect the Bose enhancement and Pauli
blocking factors that appear whenever there are excitations in the
medium. These contributions from the medium will allow us to identify similar
physical features in the case of non-linear relaxation.

\subsection{\bf Linear Relaxation: Amplitude and Perturbative
expansion}

To explore the behavior of the inflaton within the linear regime, we first use
the tadpole method to obtain the equation of motion (\ref{tad}) (see the
Appendix for details on how this method is actually applied and
reference\cite{elmfors} for an alternative implementation). The next step is
to linearize this equation in the inflaton zero mode amplitude and use this to
study the relaxational dynamics of the inflaton. We should note that this
amplitude expansion is {\it a priori} different from the standard perturbative
expansion in the relevant coupling constant. Later in this work, we will
compare the results obtained here with results obtained through a
self-consistent, non-perturbative resummation both in the coupling constants
{\em and} the field amplitude.

We arrange the initial conditions to be such that the fields $\Phi$,
$\sigma$, $\psi$ start to interact at a time that we choose $t=0$. This can be
accomplished by making the coupling ``constants'' time dependent, i.e. zero for
$t<0$ and different from zero for $t>0$. To perform the calculations, we will
need the non-equilibrium Green's functions and Feynman rules. Since the
non-equilibrium generating functionals involve a forward and backward time
contour\cite{tadpole,landsman,noneq}, the number of vertices is doubled.
Those in which
all the fields are on the forward branch (fields labeled by ($+$)) are the
usual interaction vertices, while those in which the fields are on the backward
branch (fields labeled by ($-$)) have the opposite sign. The combinatoric
factors are the same as in usual field theory. The spatial Fourier transform of
the necessary finite (initial) temperature propagators are:

\begin{itemize}

\item{Bosonic Propagators

\begin{eqnarray}
&& G_k^{++}(t,t') = G_k^{>}(t,t')\Theta(t-
t')+G_k^{<}(t,t')\Theta(t'-t)\;,
\label{timeord}  \nonumber \\
&& G_k^{--}(t,t') = G_k^{>}(t,t')\Theta(t'-
t)+G_k^{<}(t,t')\Theta(t-t')\;,
\label{antitimeord} \nonumber \\
&& G_k^{+-}(t,t')= - G_k^{<}(t,t') \label{plusmin} \;, \nonumber \\
&& G_k^{-+}(t,t')= - G_k^{>}(t,t') \label{minplus} \;,  \\
&&G_k^{>}(t,t')= i \int d^3x  \; e^{-i\vec{k}\cdot\vec{x}} \;
\langle \Phi(\vec{x},t) \Phi(\vec{0},t') \rangle \nonumber \\
&&=\frac{i}{2\omega_k}\left\{[1+n_b(\omega_k)]
e^{-i\omega_k(t-t')}+n_{b}(\omega_k) e^{i\omega_k(t-
t')}\right\}\;,
\label{ggreat} \nonumber  \\
&&G_k^{<}(t,t')= i \int d^3x \; e^{-i\vec{k}\cdot\vec{x}} \;
\langle \Phi(\vec{0},t') \Phi(\vec{x},t) \rangle \nonumber \\
&&= \frac{i}{2\omega_k}\left\{[1+n_{b}(\omega_k)]
e^{i\omega_k(t-t')}+n_{b}(\omega_k)e^{-i\omega_k(t-
t')}\right\}\;,
\label{gsmall} \nonumber  \\
&&\quad \omega_k=\sqrt{\vec{k}^2+m^2}\;,\quad\quad\quad  n_{b}(\omega_k)=
\frac{1}{e^{\beta \omega_k}-1} \;,\label{bosefactor} \nonumber
\end{eqnarray}
where $m$ is the mass of the  boson.}

\item{Fermionic Propagators (Zero chemical potential)

\begin{eqnarray}
&& S_{\vec{k}}^{++}(t,t')=S_{\vec{k}}^{>}(t,t')\Theta(t-t')
+S_{\vec{k}}^{<}(t,t')\Theta(t'-t) \;,\label{fertimeord}\nonumber\\
&& S_{\vec{k}}^{--}(t,t')=S_{\vec{k}}^{>}(t,t')\Theta(t'-t)
+S_{\vec{k}}^{<}(t,t')\Theta(t-t') \;,\label{ferantitimeord}\nonumber\\
&& S_{\vec{k}}^{+ -}(t,t')=-
S_{\vec{k}}^{<}(t,t')\;,\label{ferplusmin}\nonumber\\
&& S_{\vec{k}}^{- +}(t,t')=-
S_{\vec{k}}^{>}(t,t')\;,\label{ferminplus}\nonumber\\
&& S_{\vec{k}}^{>}(t,t')=-i\int d^3x \; e^{-i\vec{k}\cdot
\vec{x}} \; \langle \psi(\vec{x},t) \bar{\psi}(\vec{0},t')
\rangle \label{sgreatdef} \\
&&\quad\quad\quad =-\frac{i}{2\omega_k}\left[e^{-i\omega_k(t-t')}
(\not\!{k}+m_{\psi})
(1-n_{f}(\omega_k))+e^{i\omega_k(t-t')}\gamma_0
(\not{k}-m_{\psi})\gamma_0 n_{f}(\omega_k) \right]\;,
\nonumber \\
&& S_{\vec{k}}^{<}(t,t')=i\int d^3x \; e^{-i\vec{k}\cdot
\vec{x}} \;  \langle \bar{\psi}(\vec{0},t')  \psi(\vec{x},t)
\rangle \nonumber \\
&&\quad\quad\quad = \frac{i}{2\omega_k}\left[e^{-i\omega_k(t-t')}
(\not\!{k}+m_{\psi})
n_{f }(\omega_k)+e^{-i\omega_k(t-t')}\gamma_0
(\not\!{k}-m_{\psi})\gamma_0 (1-n_{f}(\omega_k)) \right]\;,
\label{slessdef}\nonumber\\
&&\quad \omega_k=\sqrt{\vec{k}^2+m_{\psi}^2}\;,\quad\quad\quad
n_{f}(\omega_k)=\frac{1}{e^{\beta \omega_k}+1}\;. \label{fermifactor}\nonumber
\end{eqnarray}}
\end{itemize}
In the linear amplitude approximation, corresponding to linear relaxation,
we find in all cases the following form of the equation of motion for the
expectation value (see Appendix)
\begin{eqnarray}
&&\ddot{\delta_{\vec{p}}}(t) + \Omega^2_{\vec{p}} \;
\delta_{\vec p}(t) +
\int_0^\infty K_{\vec{p}}(t-t') \;  \delta_{\vec{p}}(t') \;
dt'=0 \;, \label{evol}
 \label{eqnofmotion}\\
&&K_{\vec{p}}(t-t') = \Sigma_{r,\vec{p}}(t-t') \Theta(t-t')\;,
\label{retardedkernel}\nonumber
\end{eqnarray}
where we have imposed as boundary conditions that the inflaton and the other
fields are coupled at time $t=0$ but uncoupled for previous times, and
introduced
\begin{equation}
\delta_{\vec{p}}(t)=\int d^3 x \; e^{-i{\vec p}\cdot{\vec{x}}}
\;\phi(\vec{x} ,t)\,, ~~
 \Omega_{\vec{p}}^2 = \vec{p}^{\;2} + m_{\Phi}^2+\delta m(T) \; ,
\label{varios}
\end{equation}
where $\delta m(T)$ is the time independent (but temperature dependent)
contribution from tadpole diagrams. These contributions renormalize the mass
and introduce a temperature dependent effective mass and will be specified
later in each particular case. The quantity $\Sigma_{r,\vec{p}}(t-t')
\Theta(t-t')$ is the retarded self-energy. It will be computed to
dominant order in the couplings for
both fermions and bosons. Although the self-energy at finite temperature has
been computed before in the literature\cite{weldon1,keil}, we differ from
previous treatments in that we perform our calculations directly in real time,
this allows us to study real time relaxation as an initial condition problem.

Eq. (\ref{evol}) can be solved by Laplace transform. Define
\begin{equation}\nonumber
\varphi_{\vec{p}}(s)=\int_0^\infty e^{-st} \;\delta_{\vec{p}}(t) \;dt\;.
\end{equation}
Then, eq. (\ref{evol}) becomes
\begin{equation}
s^2 \varphi_{\vec{p}}(s)-s \delta_{\vec p}(0)-
\dot{\delta}_{\vec{p}}(0)
+\Omega_{\vec{p}}^2\;\varphi_{\vec p}(s)
+ \varphi_{\vec{p}}(s)\; \Sigma_{\vec{p}}(s)=0\;, \label{trflap}
\end{equation}
with $\Sigma_{\vec{p}}(s)$ the Laplace transform of
$\Sigma_{r,\vec p}(t)$.

 For computational purposes, it can
be shown that at zero temperature each graph of
$\Sigma_{\vec p}(s)$ is exactly equal to the corresponding graph
of the zero-temperature equilibrium Euclidean quantum field theory,
$s$ being the time component of the Euclidean four momentum.

In general $\Sigma_{\vec p}(s)$ can be written as a dispersion
integral in terms of the spectral density $\rho(p_o,\vec{p},T)$
\begin{equation}
\Sigma_{\vec{p}}(s)=  -\int \frac{2\, p_o\,
\rho(p_o,\vec{p},T)}{s^2+p_o^2}
dp_o \;. \label{dispersion}
\end{equation}
The imaginary part of the self-energy is found to be
\begin{eqnarray}
&& {\mbox{Im}\,} \Sigma_{\vec{p}}(s=i\omega\pm 0^+) = \pm\Sigma_{I
\vec{p}}(\omega)\;,
\nonumber \\
&& \Sigma_{I \vec{p}}(\omega)= \pi \;{\mbox{sign}\,}(\omega)
\left[\rho(|\omega|,\vec{p},T) -
\rho(-|\omega|,\vec{p},T) \right]\;. \label{imaginarypart}
\end{eqnarray}
The presence of ${\mbox{sign}\,}(\omega)$ in the above expression
characterizes the {\it{retarded}} self-energy.

Let us choose $\delta_{\vec{p}}(0)=\delta_i \; , {\dot
\delta}_{\vec{p}}(0) = 0$ for simplicity. We get from eq.
(\ref{trflap})
\begin{equation}
\varphi_{\vec p}(s)= \delta_i \;
{{s}\over{s^2+\Omega_{\vec{p}}^2
+ \Sigma_{\vec p}(s)}}\;. \label{laplatrans}
\end{equation}
The Laplace transform can be inverted through the formula
\begin{equation}
\delta_{\vec p}(t)=\int_{-i \infty+\epsilon}^ {+i
\infty+\epsilon} e^{st}
 \;\varphi_{\vec p}(s) \;
{{ds}\over{2\pi i}}\;, \label{invlap}
\end{equation}
where $\epsilon$ is a positive real constant (Bromwich
countour). Thus we need to understand the singularities of
$[s^2+\Omega_{\vec{p}}^2 + \Sigma_{\vec p}(s)]^{-1}$.
We now consider the following cases:

\begin{itemize}
\item{The inflaton potential admits spontaneous symmetry breaking (SSB),
and is only coupled to lighter {\it scalar} fields.}
\item{The inflaton is coupled to fermions only.}
\end{itemize}

We will also consider the subcases where the initial temperature is taken to be
zero, as would be appropriate in the case of evolution in the post inflationary
universe, as well as the situation where the initial temperature is non-zero,
which would be relevant to the situation of a scalar field starting
in an initial
(non-equilibrium) but thermal state and evolving out of it.

\subsection{\bf SSB with Coupling to Lighter Scalars Only}
If $m_{\Phi}^2= - \mu^2 <0 $, then the new minimum is at
$\Phi_0=\sqrt{6\mu^2 /
\lambda_{\Phi}}$, and we write
$\Phi^{\pm}(\vec{x},t)=\Phi_0+\phi(\vec{x},t)+\chi^{\pm}(\vec{x},t)$.  The
masses are now shifted to
\begin{equation}
M^2 =2\mu^2\;, \quad\quad\quad\quad
M^2_{\sigma}=m^2_{\sigma}+g^2\Phi_0^2\;.
\nonumber
\end{equation}
The contribution from the quartic inflaton self-coupling has been studied
previously\cite{reheating} thus it will not be repeated here.

The tadpole correction to the mass in eq. (\ref{varios}) is given by
\begin{eqnarray}
&& \delta M(T)= -ig \int\frac{d^3k}{(2\pi)^3}
G^{++}_{k,\sigma}(t,t)= g
\int\frac{d^3k}{(2\pi)^3}
\frac{1+2\, n_{b}(\omega_k)}{2\omega_{k}}\;,
\label{deltamass} \nonumber \\
&&  \omega_{k}  = \sqrt{\vec{k}^2+M^2_{\sigma}}\;.
\label{sigmafreq} \nonumber
\end{eqnarray}
This mass renormalization does not influence the dynamics.
The retarded self energy is found to be at order $g^2$,
\begin{equation}
K_{\vec{p}}(t-t')= 2 i  g^2 \Phi^2_0
\int\frac{d^3k}{(2\pi)^3}
\left[G^{++}_{\vec{k},\sigma}(t-t')
G^{++}_{\vec{k}+\vec{p},\sigma}(t-t')-
G^{<}_{\vec{k},\sigma}(t-t')
G^{<}_{\vec{k}+\vec{p},\sigma}(t-t') \right].
\nonumber
\end{equation}
With the non-equilibrium Green's functions defined above, we find
\begin{eqnarray}
\Sigma_{r,\vec{p}}(t-t') & = &
-2 g^2 \Phi^2_0 \int \frac{d^3k}{(2\pi)^3}\frac{1}{2
\omega_{\vec{k}}\omega_{\vec{k}+\vec{p}}}
\left\{[1+2\, n_{b}(\omega_k)]
\sin[(\omega_{\vec{k}}+\omega_{\vec{k}+\vec{p}})(t-t')]
\right. \nonumber \\
                                      &    &\left. -2\, n_{b}(\omega_k)
\sin[(\omega_{\vec{k}}-\omega_{\vec{k}+\vec{p}})(t-t')]
\right\}\;. \label{bosonselfener}
\end{eqnarray}
The Laplace transform can be written as a dispersion integral in terms of the
bosonic spectral density $\rho_b(p_o,\vec{p},T)$ (see
eq. (\ref{dispersion})). We have to one loop level:
\begin{eqnarray}
&& \rho_b(p_o,\vec{p},T)= 2g^2 \Phi_0^2 \int \frac{d^3 k}{(2\pi)^3
2\omega_{k}}\int \frac{d^3 k'}{(2\pi)^3
2\omega_{k'}}(2\pi)^3 \delta^3(\vec{p}-
\vec{k}-\vec{k}') \nonumber   \\
&& \quad\quad\times\left[\delta(p_o-\omega_{k}-\omega_{k'})
(1+\, 2\, n_{b}(\omega_k))-
\delta(p_o-\omega_{k}+\omega_{k'})\,  2\, n_{b}(\omega_k) \right]\;.
\label{disprel}\nonumber
\end{eqnarray}
The imaginary part of the self-energy is given by eq. (\ref{imaginarypart}).

We will analyze only the case of a spatially constant order parameter
corresponding to $\vec{p}=0$ because we will later compare with the case of
non-linear relaxation which we only study for a homogeneous (translational
invariant) expectation value. In this case the spectral density can be written
as
\begin{equation}
\rho_b(p_o,\vec{0},T) =
\left[1+2\, n_{b}\left(\frac{p_o}{2}\right)
\right] \rho_b(p_o,\vec{0},0)\;. \label{finitetspecdens}
\end{equation}
The spectral density $ \rho_b(p_o,\vec{0},0)$ is a Lorentz scalar and is
proportional to the decay rate of the boson $\Phi$ into two $\sigma$
particles. It is a straightforward exercise to find
\begin{equation}
\rho_b(p_o,\vec{0},0) = \frac{g^2 \Phi_0^2}{8 \pi^2}
\left[1- \frac{4 M^2_{\sigma}}{p_o^2} \right]^{\frac{1}{2}}
\Theta\left( p^2_o -4 M^2_{\sigma}\right) \;.\label{specdens}
\end{equation}
Clearly $ \Sigma_{\vec{0}}(s)$ has a logarithmic divergence, independent of $s$
and $T$. We choose to subtract this divergence at $s=0$ and absorb the
subtraction into a further temperature dependent renormalization of $M$.  In
order not to clutter notation, from now on we will refer to $M$ as the fully
renormalized mass including the above subtraction, and to $\Sigma(s) \equiv
\Sigma_{\vec{0}}(s) - \Sigma_{\vec{0}}(0)$. From equations
(\ref{dispersion}, \ref{finitetspecdens}, \ref{specdens}) we find that the
self-energy has an imaginary part above the two $\sigma$-particles threshold
given by eq. (\ref{imaginarypart}) with
\begin{equation}
\Sigma_I(\omega)  =  \frac{g^2 \Phi^2_0}{8 \pi}
\left[1- \frac{4 M^2_{\sigma}}{\omega^2}\right]^{\frac{1}{2}}
\left[1+2\,n_b\left(\frac{\omega}{2}\right)\right]
\Theta \left(\omega^2 -4 M^2_{\sigma}\right) \mbox{ sign }(\omega) \;.
\label{imagpart}
\end{equation}
This expression is recognized as the imaginary part of the retarded self-energy
(determined by the sign($\omega$)) and shows the usual Bose enhancement
factor\cite{weldon2}.

\subsubsection{\bf{Zero Temperature}}

In this case from \ref{bosonselfener} we find
\begin{equation}
\Sigma(s)=\frac{g^2 \Phi_0^2}{4 \pi^2}
\left(\sqrt{1+\frac{4 M^2_{\sigma}}{s^2}}
\;{\mbox{ArgTh}\,}\frac{1}{\sqrt{1+\frac{4
M^2_{\sigma}}{s^2}}}-1\right)\;.
\label{Sigun}
\end{equation}
In order to compute the inverse Laplace transform through eq. (\ref{invlap})
we must first study the analytic structure of $\varphi(s)$ in the $s$-plane.
$\varphi(s)$ has poles at the zeroes of
\begin{equation}
s^2+M^2+\Sigma(s)=0\;. \label{ecpol}
\end{equation}
These correspond to $\Phi$-one-particle states with the mass including
one-loop radiative corrections.
At zeroth order the poles are purely imaginary
\begin{equation}
s_\pm=\pm iM\;.\nonumber
\end{equation}
To find the one-loop correction, we set
\begin{equation}
s_+=iM+ r\nonumber
\end{equation}
and similarly for $s_-$. Inserting this in eq. (\ref{ecpol}) yields to order
$g^2$
\begin{equation}
2i  Mr+\Sigma(iM)=0\;.\nonumber
\end{equation}
That is,
\begin{equation}
r = i \frac{\Sigma(iM)}{2M}\;.  \label{polo}
\end{equation}
When $M<2M_{\sigma}$, $\Sigma(iM)$ is real (see eq. (\ref{Sigun})) and
eq. (\ref{polo}) gives a real correction to the $\Phi$ mass
\begin{equation}
s_\pm = \pm iM_0 \equiv \pm i \left[M  +{{g^2 \Phi^2_0}\over{8\pi^2 M}}
\left(\sqrt{{{4 M^2_{\sigma}}\over{M^2}}-1}
\;\;{\mbox{ArgTh}\,}{1\over{\sqrt{{{4 M^2_{\sigma}}\over{ M^2}}-1}}}-
1\right)\right]\;.  
\nonumber
\end{equation}
The Laplace transform $\varphi(s)$ also exibits a cut along the
imaginary axis starting at $s = i\omega = \pm 2iM_{\sigma}$.
For $s$ in the first Riemann sheet (physical sheet) we
obtain
\begin{equation}
\Sigma_{\rm physical}(i\omega\pm 0^+)=\Sigma_R(\omega)\pm
i\Sigma_I(\omega)\;,
\;\;\;\;\;\;\;\;\;\;\; \omega > 2 M_{\sigma}
\nonumber\;,
\end{equation}
with $\Sigma_R$ and $\Sigma_I$ both real and given by
\begin{equation}
\Sigma_R(\omega)
={{g^2\Phi_0^2}\over{4\pi^2}}\left(\sqrt{1-{{4 M^2_{\sigma}}\over{\omega^2}}}
\;\;{\mbox{ArgTh}\,}\sqrt{1-{{4 M^2_{\sigma}}\over{\omega^2}}} -1\right)\,,
\; \;\quad \Sigma_I(\omega)={{g^2 \Phi^2_0}\over{8\pi}}
\sqrt{1-{{4 M^2_{\sigma}}\over{\omega^2}}}>0\;.
\nonumber
\end{equation}
We can now proceed to compute the inverse Laplace transform
(\ref{invlap}) by deforming the contour.
\begin{equation}
\delta(t)=\frac{\delta_i \cos M_0 t}
{1-\frac{\partial\Sigma(iM)}{\partial M^2}}
+{{2\delta_i }\over{\pi}} \int_{2M_{\sigma}}^{\infty}
{{\omega\Sigma_I(\omega) \cos\omega
t\;d\omega}\over{[\omega^2-M^2-
\Sigma_R(\omega)]^2+ \Sigma_I(\omega)^2}}\;. \label{stable}
\end{equation}
For large time $M_{\sigma}t \gg 1$ the integral over the cut is dominated
by the endpoint $\omega=2M_{\sigma}$ and goes to zero as
\begin{equation}
\delta_{\rm cut}(t)\simeq
{{\delta_i \; \sqrt{\pi}\; M^2_{\sigma} \; g^2\Phi_0^2}\over{4\pi^2(M^2-
4M^2_{\sigma}+ {{g^2\Phi_0^2}\over{4\pi^2}})^2}}
{{\cos(2M_{\sigma}t+{{3\pi}\over{4}})}\over{(M_{\sigma}t)^{3
\over 2}}}\;. 
\nonumber
\end{equation}
The $t^{-3/2}$ is completely determined by the behavior of the spectral
density at threshold.

The situation changes drastically for $M>2M_{\sigma}$. In such case the
$\Phi$-particle is unstable and thus $\Sigma(iM)$ becomes complex and its value
depends from which side of the cut we approach the imaginary axis. Now the
solution for the pole will be complex and $r$ will acquire a {\em real}
part. Due to the discontinuity in $\Sigma_I$ across the two-particle cut,
eq. (\ref{polo}) must be written carefully as
\begin{equation}
r=\frac{i}{2M} \; \Sigma(iM+\mbox{Re}\,[r])\;,\nonumber
\end{equation}
where the (small) real correction inside the argument will determine on which
side of the cut the solution resides.  Then the real part of $r$ should satisfy
\begin{equation}
\mbox{Re}\, r = -{{1} \over {2M}}\, \mbox{Im}\,\Sigma(iM+ 0^+ \;
{\mbox{sign}\,}(\mbox{Re}\,[r]))\;.
\label{anch}
\end{equation}
{F}rom eq. (\ref{imagpart}) we see that $\mbox{Im}\,\Sigma_{\rm physical}$
is negative for ${\mbox{sign}\,}(\mbox{Re}\,[r])<0$ and positive for
${\mbox{sign}\,}(\mbox{Re}\,[r])>0$.
Therefore eq. (\ref{anch}) {\it has no solution} in the physical sheet.

The analytic continuation of $\Sigma$ into the second Riemann sheet is such
that\cite{Smat}
\begin{equation}\nonumber
\Sigma^{II}(i\omega\pm 0^+)=\Sigma_R(\omega)\mp
i\Sigma_I(\omega)\,,
\;\;\;\;\;\;\;\;\;\;\;\omega>2M_{\sigma}
\end{equation}
and we find the solution
\begin{eqnarray}
&&\mbox{Re}\, r=-\frac{1}{2M}\Sigma_I(M)=
-{g^2\Phi^2_0\over{16\pi M}}\sqrt{1-{{4
M^2_{\sigma}}\over{M^2}}}<0\;, \nonumber \\
&&\mbox{Im}\, r=\frac{1}{2M}\Sigma_R(M)
={{g^2\Phi^2_0}\over{8\pi^2 M}}\left(\sqrt{1-{{4
M^2_{\sigma}}\over{M^2}}}
\;\;{\mbox{ArgTh}\,}\sqrt{1-{{4 M^2_{\sigma}}\over{M^2}}}
-1\right)\;. \label{polins}\nonumber
\end{eqnarray}
For $M\gg M_{\sigma}$,
\begin{equation}
\mbox{Im}\, r={ g^2\Phi^2_0\over{8\pi^2 M}}\;\left(
\log{{M}\over{M_{\sigma}}}-1 \right)\;.\nonumber
\end{equation}
$|\mbox{Re}\,[r]|$ coincides with the decay rate
$\Phi \to 2 \sigma $ (as it must be)
which is the rate per unit time to produce $\sigma$ particles.
The poles $s_{\pm}$ move  off into the  second sheet when $M$ becomes
larger than $2M_{\sigma}$ as expected\cite{Smat}.

Thus when we compute the inverse Laplace transform
(\ref{invlap}) in the unstable case
we are left with the integral over the cuts since both poles
are in the second Riemann sheet and we find
\begin{equation}
\delta(t)=\frac{2 \delta_i}{\pi} \;  \int_{2M_{\sigma}}^\infty
{{\omega\Sigma_I(\omega) \cos\omega t\,d\omega}\over{[\omega^2-M^2-
\Sigma_R(\omega)]^2+ \Sigma_I(\omega)^2}}\;.
\nonumber
\end{equation}
Since now $M$ is inside the integration region, for weak coupling there is a
narrow resonance at $\omega\simeq M$. Thus for weak coupling it takes
Breit-Wigner
form and we find to a very good approximation,
\begin{equation}
\delta(t)\simeq\delta_i\, A\; e^{-{\Gamma t/2}}\;
\cos(Mt+\alpha)
\;,\;\;\;\;\quad\quad\quad\Gamma \ll M \;,\nonumber
\end{equation}
where
\begin{equation}
A=1+ {{\partial\Sigma_R(M)}\over{\partial M^2}}
\;,\quad\quad
\Gamma =  {g^2\Phi^2_0\over{8\pi M}}\sqrt{1-{{4
M^2_{\sigma}}\over{M^2}}}  \;,
\quad\quad
\alpha= -{{\partial\Sigma_I(M)}\over{\partial M^2}}\;.\nonumber
\end{equation}
For $M\gg M_\sigma$,
\begin{equation}
A=1+{{g^2\Phi^2_0}\over{8 \pi^2 M^2}}+
{\cal O}\left(\left[\frac{M_\sigma}{M}\right]^2\right)\;,
\;\;\;\;\;\;\;\;\;\;\;
\alpha={{g^2 \Phi^2_0 M_\sigma^2}\over{4 \pi M^4}}
+{\cal O}\left(\left[\frac{M_\sigma}{M}\right]^4\right)\;.
\nonumber
\end{equation}
The Breit-Wigner approximation, however, is valid only for times
$\leq\Gamma^{-1}\ln(\Gamma / M_\sigma)$;
for longer times the fall off is with a power law $t^{-3/2}$
determined by the spectral density at threshold as before.

\subsubsection{\bf{Non-Zero Temperature}}
The physical mass gets a finite temperature dependent correction from the
tadpole contribution $\delta M(T)$ given by
\begin{equation}
\delta M(T) =g \int_0^\infty {{dk}\over{2\pi^2}}
{{k^2}\over{\sqrt{k^2+M_\sigma^2}}}
{{1}\over{e^{\beta\sqrt{k^2+M_\sigma^2}}-1}}\;.\nonumber
\end{equation}
For small $\beta$ ($T \gg M_\sigma$) this correction takes the form
\begin{equation}\nonumber
\delta M(T)=g\left\{ \frac{T}{12} -
\frac{M_\sigma T}{4\pi}
+\frac{M_\sigma^2}{8\pi}\log(\frac{T}{M_\sigma})+{\cal O}(T^0) \right\}\;.
\end{equation}
We will assume that $gT^2 \ll M^2$ so that we are in the perturbative
regime, since otherwise hard thermal loops
must be resummed\cite{pisarski,braaten},
 a task beyond the scope of this article.

The imaginary part of the self-energy is given in eq. (\ref{imagpart}) and
the real part can be obtained from the dispersion integral (\ref{dispersion})
using (\ref{finitetspecdens}, \ref{specdens}). The real part of the self-energy
is difficult to compute for arbitrary temperature, and we just quote its large
$T$ behavior:
\begin{equation}
\Sigma(s,\beta)=g^2 \Phi^2_0 \frac{M_\sigma T}{\pi s^2}
\left[1-\sqrt{1+\frac{s^2}{4M_\sigma^2}}\right]
+{\cal{O}}(T^0)\;.
\nonumber
\end{equation}
It is interesting to compute $\Sigma(t,\beta)$ in configuration space
for large $T$. We have by inverse Laplace transform
\begin{equation}
\Sigma(t,\beta) =\int_{-i \infty+\epsilon}^ {+i
\infty+\epsilon} e^{st}
\;\Sigma(s,\beta) \; {{ds}\over{2\pi i}}\;. 
\nonumber
\end{equation}
Upon contour deformation we find
\begin{equation}
\Sigma(t,\beta)=-{{g^2 \Phi^2_0 T}\over {\pi^2}} \int_{1}^{\infty}
{{dx}\over x}\;\sqrt{x^2-1}\;\sin(2M_\sigma xt)\;.\nonumber
\end{equation}
This function is obviously {\bf not} concentrated at $t=0$.
It can be related to a $J_1$ Bessel function as
\begin{equation}
\Sigma(t,\beta)= -{{g^2 \Phi^2_0 T}\over {2 \pi}}
\left[1-{{4M_\sigma}\over{\pi}}\, t -2\,M_\sigma \int_0^t dx\;
\left(\frac{t}{x} -1 \right)\; J_1(2\,M_\sigma x) \right]
\end{equation}
We find for small and for large $t$
\begin{eqnarray}
\Sigma(t,\beta) &\buildrel {t\to 0} \over =& -{{g^2 \Phi^2_0 T}\over {2 \pi}}
\left[1-{{4M_\sigma}\over{\pi}}\, t+O(t^2)\right] \nonumber \\  \nonumber \\
\Sigma(t,\beta) &\buildrel {t\to \infty} \over =& {{g^2 \Phi^2_0
T}\over{2\pi^2}} {{\sin(2M_\sigma t - \pi/4)}\over {(M_\sigma t)^{3/2}}}
\left[1+O(t^{-1}) \right]\;.\nonumber
\end{eqnarray}
The behaviour of the kernel $\Sigma(t,\beta)$ shows that it {\bf
cannot} be approximated by a phenomenological term $\Gamma\, {d \over
{dt}}$ even for high $T$. (As shown in ref.\cite{reheating} this is
not the case either for $T=0$).

We can now repeat the analysis of the previous section for both cases
$M<2M_{\sigma}$ and $M>2M_{\sigma}$. In the first case, the one particle pole
is below the two $\sigma$-particles threshold with a (small) finite temperature
correction since we are restricted to the perturbative regime in which $gT^2
\ll M$. The long time behavior will be oscillatory with the frequency
corresponding to the one particle pole plus long-time power law tails similar
to the zero temperature case.  The second case is more interesting, since now
the scalar ``inflaton'' is unstable, and the pole moves off into the second
Riemann sheet. In the physical sheet there is a resonance with a finite
temperature width given by
\begin{equation}
\Gamma(T)=\Gamma(0)\left[1+2\, n_b\left(\frac{M}{2}\right)\right]\,, ~~
\quad\quad\quad
\Gamma(0)={g^2\Phi^2_0\over{8\pi M}}\sqrt{1-{{4
M^2_{\sigma}}\over{M^2}}}\;.
\nonumber
\end{equation}
The Bose enhancement factor increases the rate and therefore enhances
dissipation via the production of particles. Although this factor arises from
the thermal distribution, we expect in general that whenever there are bosonic
excitations present, the relaxation rate will be enhanced as a consequence of
bose statistics, independently of whether these excitations are thermally
distributed.

\subsection{\bf  Unbroken symmetry with coupling to light scalars}
In the unbroken symmetry case $M_{\Phi}$ is above all thresholds and
hence $\Phi$ is a stable particle. The order parameter is again given
by a formula like eq. (\ref{stable}) except that now the integration
starts at $\omega = M+2 M_\sigma$.

The first perturbative contribution to
the kernel $\Sigma(t-t')$ in the inflaton equation of motion (\ref{evol})
are now the two loop order graphs usually
called ``setting sun''. Since these graphs are quite complicated,
we only perform the two loop computation at zero temperature, for
which we can exploit the relationship with usual Euclidean field theory.
Even with this simplification the computation is hard, and we
limit ourselves to the evaluation of the imaginary part of the retarded
self-energy near the branch point. We have seen
below  eq. (\ref{stable}) how this
actually determines the long time behaviour of the order parameter.
In general, if $\Sigma_I(\omega\to\omega_{\rm{threshold}})$ vanishes as
$(\omega-\omega_{\rm{threshold}})^\alpha$, then
$\delta_{{\mbox{\footnotesize{cut}}}}(t)$ decays as
$t^{-1-\alpha}$ for large times.

We find for the two loops self-energy
\begin{equation}
\Sigma_I(\omega\to M+2 M_\sigma)\simeq
\frac{2 g^2 \pi^2}{(4\pi)^4}
\frac{M_\sigma \sqrt{M}}{(M+2 M_\sigma)^{7/2}}
[\omega^2-(M+2 M_\sigma)^2]^2\;.\nonumber
\end{equation}
This yields for $\delta_{{\mbox{\footnotesize{cut}}}}(t)$ a power decay
$t^{-3}$ for long times.

For massless $\sigma$ we find
\begin{equation}
\Sigma_I(\omega\to M)\simeq
\frac{g^2 \pi}{3 M^4 (4\pi)^4}(\omega^2-M^2)^3\;,\nonumber
\end{equation}
yielding a $\delta_{{\mbox{\footnotesize{cut}}}}(t)$ decaying as
$t^{-4}$ for long times.
In summary, in the unbroken symmetry case $\delta(t)$ is given by the
one-particle term oscillating with frequency $M_0$ plus the power damped
cut contribution  $\delta_{{\mbox{\footnotesize{cut}}}}(t)$.

\subsection{\bf Inflaton coupled to Fermions}
Fermions can be treated similarly in the broken and unbroken symmetry phase.
To treat both cases on equal footing we now define $M$ as mass of the inflaton
(scalar) field and $m$ as the fermion mass in either case. Using the Feynman
rules described in the first section, we obtain to one-loop level the kernel
\begin{eqnarray}
K_{\vec{p}}(t-t')           & =  & i y^2 \int \frac{d^3 k}{(2\pi)^3}
 {\mbox{Tr\,}} \left[ iS^{++}_{\vec{k}}(t,t')
iS^{++}_{\vec{k}-\vec{p}}(t',t) - iS^{<}_{\vec{k}}(t,t')
 iS^{>}_{\vec{k}-\vec{p}}(t',t)\right]\;, \label{fermikernel} \nonumber \\
\Sigma_{r,\vec{p}}(t-t') & = & i y^2 \int \frac{d^3 k}{(2\pi)^3}
 {\mbox{Tr\,}} \left[ iS^{>}_{\vec{k}}(t,t')
iS^{<}_{\vec{k}-\vec{p}}(t',t) - iS^{<}_{\vec{k}}(t,t')
 iS^{>}_{\vec{k}-\vec{p}}(t',t)\right]\;. \label{retfermiker}\nonumber
\end{eqnarray}
We now  concentrate on the homogeneous case $\vec{p}=0$. Using the fermionic
Green's functions given in the first section it is straightforward to
find the fermionic spectral density to be used in eq.(\ref{dispersion})
\begin{eqnarray}
\rho_f(p_o,T)  & = & \frac{y^2}{8\pi^2}  \; p^2_o
 \left[1-2n_f\left(\frac{p_o}{2}\right)\right]
\left[1- \frac{4 m^2}{p_o^2} \right]^{\frac{3}{2}}
 \Theta\left( p^2_o -4 m^2\right)\;.\label{ferspecdens}\nonumber
\end{eqnarray}
 From eq.(\ref{imaginarypart}), the imaginary part of the self energy is then
\begin{eqnarray}
&&\Sigma_I(\omega)  =  \frac{y^2 }{8 \pi}\omega^2
\left[1- \frac{4 m^2}{\omega^2}\right]^{\frac{3}{2}}\;
\left[1-2n_f\left(\frac{\omega}{2}\right)\right] \;
\Theta \left(\omega^2-4 m^2 \right) \;.
\label{ferimagpart}\nonumber
\end{eqnarray}
The finite temperature factor reflects the Pauli blocking term\cite{weldon2}.
It is clear that the zero temperature part of $\Sigma(s)$ diverges
quadratically and that two subtractions are needed. The first one is
independent of $s$ and contributes a (quadratically divergent) mass
renormalization. The second one is logarithmic divergent and consists of an
$s$ independent term that adds to the mass renormalization and another
proportional to $s^2$ that will be absorbed in wave function renormalization.
We choose to subtract at zero temperature and at an arbitrary scale
$\kappa$. The Laplace transform for the zero momentum component of the equation
of motion (\ref{laplatrans}) becomes
\begin{equation}
\varphi(s) = \frac{\delta_i \; s}{  s^2 \left(1+
\frac{y^2}{4\pi^2} \ln \frac{\Lambda}{\kappa}\right)
+ M^2_{1R}(T)+
y^2 \; \Pi \left(s,T, \kappa \right) }\;,
\nonumber
\end{equation}
where $M_{1R}^2$ contains the mass renormalization
and $\Pi$ is the twice subtracted kernel,
which depends on the renormalization scale. Defining
\begin{eqnarray}
  Z^{-1}_{\phi}(\kappa)& = &  1+\frac{y^2}{4\pi^2}\ln
\frac{\Lambda}{\kappa}\;,
 \label{wavefunctionren}\nonumber \\
y_R (\kappa)                 & = & Z^{\frac{1}{2}}_{\phi} \;  y\;,
\label{couplingren}\nonumber \\
M_R(T,\kappa)             & = & Z^{\frac{1}{2}}_{\phi} \;  M_{1R}(T)\;,
\label{finalrenormass}\nonumber \\
\varphi_R(s,\kappa)            & = & Z^{-1}_{\phi}  \; \varphi(s)\;,
\label{renormafield}\nonumber
\end{eqnarray}
we finally obtain the renormalized Laplace transform
\begin{equation}
\varphi_R(s,\kappa) = \frac{\delta_i \;  s}{ s^2 + M^2_R (T,\kappa)+
y^2_R (\kappa)  \; \Pi \left(s,T,\kappa\right)} \;.\label{renofermiequ}
\end{equation}
The inverse Laplace transform is obtained as in equation (\ref{invlap}). The
result will be the function $\delta_{R,\vec{0}}(t,\kappa)$ which is not a
renormalization group invariant. It is clear, however, from the renormalization
prescriptions described above that ratios of the amplitude at different times,
such as $\delta_{R,\vec{0}}(t,\kappa)/ \delta_{R,\vec{0}}(0,\kappa)$ are
renormalization group invariant. Now the analysis can proceed as in the
previous section.  To obtain the inverse Laplace transform we must recognize
the singularities in (\ref{renofermiequ}).  If $M_R < 2 m$ ($m$ is the fermion
mass in the loop), there is a one particle pole (with strength different from
one because we decided to renormalize off-shell) and a two-fermion cut at $ s^2
= -4m^2$. At long times the amplitude will oscillate with an oscillation period
given by the position of the pole, which is perturbatively close to $M_R$. The
contribution of the cut falls-off at long times as $t^{-5/2}$ and is determined
by the behavior of the spectral density near the two-fermion threshold.

More interesting is the case in which $M_R > 2m$. As in the bosonic case, the
pole moves off the physical sheet into the second sheet. In the physical sheet
the spectral density at weak coupling features a sharp peak at $M_R
+{\cal{O}}(y^2)$ with width
\begin{equation}
\Gamma(T)  =  \Gamma(0) \left[1-2n_f\left(\frac{M_R}{2}\right)\right]\;,~~
\quad\quad\quad\Gamma(0)  =  \frac{y^2 }{8 \pi}M_R
\left[1- \frac{4 m^2}{M^2_R}\right]^{\frac{3}{2}}\;. \label{widthfer}
\end{equation}
A Breit-Wigner approximation predicts exponential relaxation but eventually
at long times (an estimate similar to the bosonic case) a power law relaxation
$t^{-5/2}$ ensues, completely determined by the spectral density at threshold.

Pauli blocking makes the resonance narrower and the lifetime of the decaying
particle longer.  The interpretation of this phenomenon is simple. In the
thermal bath, fermionic excited states are filled with the Fermi-Dirac
distribution. In order for the scalar field to decay, it must create a
fermion-antifermion pair. However, at finite temperature, the available states
are already filled with thermal excitations and because of the Pauli exclusion
principle, are not available. At infinite temperature (and zero chemical
potential), each fermion and antifermion state are populated with occupation
$1/2$ per spin degree of freedom; in this limit the decay rate goes to zero,
the lifetime to infinity and the bosonic particle simply cannot decay because
there are no states available to decay into. Even at zero temperature, but in a
situation in which excited states are occupied, dissipative processes mediated
by the production of fermion-antifermion pairs will be hindered by the Pauli
exclusion principle, since states will already be occupied and no longer
available in the particle production process. In highly excited states, we
expect damping via production of fermion pairs to be strongly suppressed by
Pauli blocking. This phenomenon has been seen numerically in the case of
fermion pair production in strong electric fields\cite{kluger} and will be
seen numerically in the non-linear relaxation case later.

\subsection{Thermalization or Relaxation?}

{F}rom the analysis presented above, the following conclusions
for the linear regime become very clear:
\begin{enumerate}
\item{The imaginary part of the self-energy only determines a {\em relaxation
rate} in the case of a {\bf resonance}, that is when there is an imaginary part
on-shell for the external particle and the (quasi) particle pole moves off the
physical sheet into the second (unphysical) sheet. In this case a Breit-Wigner
approximation describes the exponential relaxation for a long time, but
eventually the amplitude falls-off with a power law in time, with the power
determined by the behavior of the spectral density at threshold. When the
on-shell pole is below the two-particle threshold, the imaginary part does not
translate to a damping rate. Relaxation is described by a power law with an
asymptotic behavior completely determined by the position and residue of the
pole.}

\item{Even in the case of a resonance and exponential damping, the
``damping rate'' $\Gamma$ describes exponential relaxation for the {\em
expectation value} of the scalar field. The issue of thermalization is
completely different. Thermalization corresponds to the time evolution of the
(quasi) particle distribution function towards a thermal distribution which,
in principle, has nothing to do with the relaxation of the expectation value
of the field.}
\end{enumerate}

An alternative way to look at thermalization is as a process of momentum and
energy transfer. Thus, the thermalization rate should be identified with the
rate of energy and momentum transfer which is not necessarily related to the
relaxation rate of the expectation value of the field. In order to understand
thermalization, a collisional (quantum) Boltzmann equation must be set
up. Although in the Born approximation the collision term includes the
scattering cross section that is related to the decay rate, the solution to the
Boltzmann equation implies a resummation that is quite different from the
resummation of the Dyson series for the propagator. If the initial distribution
is very far from thermal and many collisions are necessary for thermalization,
the relaxation and thermalization time scales may be widely different and in
principle unrelated. Thus we insist that the ``damping rate'' obtained from
the imaginary part of the self-energy {\em on shell} must be interpreted as the
relaxation rate for the expectation value of the scalar field and in principle
{\em not} with the thermalization rate of the particle distribution
function. Moreover, this relaxation rate holds in the linear regime
(small field amplitude).

\subsection{The Reheating Temperature:}

We can use the results obtained above to provide an
estimate of the reheating temperature
based on single particle decay and valid
{\bf only in the linear regime}\cite{linde,rocky}.
 Furthermore, since the
calculations of the relaxation and decay rate were performed in flat
Minkowski space these
will only be useful for an estimate of the reheating temperature whenever
$\Gamma \gg H$ with
$H$ being the expansion rate of the universe. Since the calculation of the
rate even in flat space
relies on a ``sharp resonance approximation'' (Breit Wigner) this implies
that  the lifetime of the
particle is much longer than its oscillation period. This translates into
the constraint
$m_{\phi}\gg  \Gamma \gg H$. The inflaton oscillates coherently  many times
around the
minimum of the potential before its oscillation amplitude diminishes
appreciably both as a result
of decay into lighter particles and because of the red-shift of the energy.

The Friedman equation for the rate of expansion is
\begin{equation}
H^2(t)= \frac{8\pi}{3 M^2_{Pl}}\rho(t) \label{hequation}
\end{equation}
with $\rho(t)$ the matter energy density.

After the reheating stage is completed it is expected that the light
produced particles will be in
the form of radiation (ultrarelativistic), thus $\rho(t) \approx T^4_{reh}$.
During the period
in which the relaxation rate (or rate of particle production) is
$\Gamma < H(t)$ the produced
particles will be red-shifted in energy and will not contribute to
a radiation dominated universe.
Only when the expansion rate becomes smaller than the rate of particle
production is when the
produced particles will remain as a radiation dominated component.
Thus the important time
scale for which the produced particles will remain as a radiation
dominated component in the
universe is that for which $H(t_{f})=\Gamma$. At this time
\begin{equation}
\rho(t_f) = \frac{3 M^2_{Pl}}{8 \pi}H^2(t_f) \approx T^4_{reh} \label{estimate}
\end{equation}
leading to the estimate of the reheating temperature within the linear regime:

\begin{equation}
T_{reh} \approx \sqrt{M_{Pl}\Gamma} \label{treh}
\end{equation}

\subsection{The Polonyi problem:}

One area of current interest in which this work may have some implications is
that of the so-called ``post-modern Polonyi problem'' concerning flat
directions for some of the moduli fields in string
theories\cite{moduli,banks2,lyth,randall}. These
are fields with pertubatively flat directions whose degeneracy is lifted by
non-perturbative effects.
Their masses $m_{\phi} \approx 10^2-10^3 \mbox{ Gev }$
with vacuum expectation values of order $10^{11}-10^{12} \mbox{ Gev }$ and
 couplings to normal matter are of order
$g \approx m_{\phi}/M_{Pl} \approx 10^{-17}$.
These properties allow the
energy density in these fields to dominate that of the radiation in the
universe until times well after nucleosynthesis.
These extremely weak couplings determine that the decay rate of
these fields is of order
\begin{equation}
\Gamma \approx \frac{m^3_{\phi}}{M^2_{Pl}} \label{polotem}
\end{equation}

A calculation of the reheating temperature
based on the ``old'' scenario of one particle decay gives
$T_{reh} \approx 0.01 - 0.1 \mbox{ Mev }$, well below the
nucleosynthesis scale.

\section{\bf Non-Linear Relaxation}
In this section we study the equation of motion for a homogeneous order
parameter beyond the linear approximation. This is achieved as
follows\cite{reheating}. We write the scalar inflaton field as
\begin{equation}
\Phi^{\pm}(\vec{x},t)= \phi(t)+\chi^{\pm}(\vec{x},t)\;,
\nonumber
\end{equation}
with $\phi(t)$ the expectation value in the non-equilibrium density matrix and
$<\chi^{\pm}(\vec{x},t)>=0$, we consider that $<\sigma^{\pm}>=0$ (this is a
consistent assumption).  The non-equilibrium path integral requires the
Lagrangian density
\begin{eqnarray}
&&{\cal{L}}(\phi; \chi^+; \sigma^+;  \chi^-; \sigma^-) =
 {\cal{L}}(\phi; \chi^+; \sigma^+)  -  {\cal{L}}(\phi; \chi^-;
\sigma^-)\;,   \nonumber \\
&&{\cal{L}}(\phi; \chi^+; \sigma^+) =
 \frac{1}{2}\left[ (\partial_{\mu}
\chi^+)^2-M^2(t)\, (\chi^+)^2\right]-\chi^+
\left[\ddot{\phi} +m^2_{\Phi}\, \phi +\frac{\lambda}{6}\, \phi^3\right]-
\frac{\lambda}{6}\phi\,  (\chi^+)^3- \frac{\lambda}{4!}\, (\chi^+)^4
\nonumber \\
&& +\frac{1}{2}\left[ (\partial_{\mu}
\sigma^+)^2-m^2(t)\, (\sigma^+)^2\right]-
g\, \phi \, \chi^+ \, (\sigma^+)^2-\frac{g}{2}\, (\chi^+)^2\,
(\sigma^+)^2 + {\bar{\psi}^+} (i {\not{\partial}} -m_{\psi}(t)
 -y \chi^+  ) \psi^+ \;, \label{plusminuslag} \nonumber \\
&& M^2(t) = m^2_{\Phi}+\frac{\lambda}{2}\, \phi^2(t) \;, \; ~~
m^2(t) = m^2_{\sigma}+g\, \phi^2(t)\;, \; ~~
m_{\psi}(t)=m_{\psi}+y\, \phi(t)\;. \label{massesoft}  \nonumber
\end{eqnarray}
The difference with the linear relaxation case is that we now incorporate
$\phi(t)$ in the definition of the time dependent masses. The equations of
motion are obtained as in the previous sections, by treating the {\it linear},
cubic and quartic terms as perturbations. The necessary Green's functions are
constructed from the homogeneous solutions of the quadratic forms. We again
treat the cases where the inflaton is coupled to scalars or fermions
separately.

\subsection{\bf Inflaton Coupled to Scalars Only}
The Green's functions for the scalars are obtained from the mode equations
that solve
\begin{eqnarray}
&&\left[\frac{d^2}{dt^2}+\vec{k}^2+M^2(t)\right] U_k(t)=0\;,
\label{umodes}  \\
&& U_k(0)=1\;, ~~ \dot{U}_k(0)= -i W_k= -i \sqrt{\vec{k}^2+M^2(0)}\;,
\nonumber \\
&&\left[\frac{d^2}{dt^2}+\vec{k}^2+m^2(t)\right] V_k(t)=0\;,
\label{vmodes} \\
&& V_k(0)=1\;, ~~ \dot{V}_k(0)= -i\,  w_k= -i \sqrt{\vec{k}^2+m^2(0)}\;,
\nonumber
\end{eqnarray}
The initial conditions on the mode functions (\ref{umodes}, \ref{vmodes})
correspond to positive frequency solutions at the initial time.
In terms of these mode functions, the Green's functions
are\cite{reheating,tadpole}
\begin{eqnarray}
&& G_{\chi,k}^>(t,t') = \frac{i}{2W_k}\left[(1+n_b(W_k))\; U_k(t)U_k^*(t')+
n_b(W_k)\; U^*_k(t)U_k(t')\right]\;, \label{ggreaterchi}  \nonumber \\
&& G_{\chi,k}^<(t,t') = G_{\chi,k}^>(t',t)\;, \label{glesserchi}  \nonumber \\
&& G_{\sigma,k}^>(t,t') = \frac{i}{2w_k}\left[(1+n_b(w_k))\; V_k(t)V_k^*(t')+
n_b(w_k)\; V^*_k(t)V_k(t')\right]\;, \label{ggreatersigma} \nonumber \\
&& G_{\sigma,k}^<(t,t') = G_{\sigma,k}^>(t',t) \label{glessersigma} \nonumber
\end{eqnarray}
and the rest of the Green's functions are given by the relations in equations
(\ref{minplus}). The Green's
functions above correspond to the situation in which the initial density matrix
is in equilibrium for the positive and negative frequency modes at time
$t=0$. This is determined by the initial conditions on the mode functions above
at an initial temperature $T=1/\beta$. However, for the rest of the analysis we
will take $T=0$. We see that
\begin{eqnarray}
&& <\chi^2(\vec{x},t)> = \int \frac{d^3k}{(2\pi)^3} \frac{|U_k(t)|^2}{2W_k}
\coth \left[\frac{\beta W_k}{2}\right]\;, \label{flucchi} \nonumber \\
&& <\sigma^2(\vec{x},t)> = \int \frac{d^3k}{(2\pi)^3} \frac{|V_k(t)|^2}{2w_k}
\coth \left[\frac{\beta w_k}{2}\right]\;. \label{flucsigma} \nonumber
\end{eqnarray}
Finally the equation of motion to one-loop order for the expectation value is
\begin{equation}
\ddot{\phi}(t)+ m^2_{\Phi}\; \phi(t)+ \frac{\lambda}{6}\; \phi^3(t)+
\frac{\lambda}{2}\; \phi(t) <\chi^2(\vec{x},t)> + g \; \phi(t)\;
<\sigma^2(\vec{x},t)> =0
\label{onelupeqn}
\end{equation}
If one wants to solve the equation of motion (\ref{onelupeqn}) to order
($\hbar$) one would expand $\phi$ in a power series in $\hbar$ and only keep
the zeroth order term in the mode equations (\ref{umodes}, \ref{vmodes}).
As was observed previously\cite{reheating}, such an expansion will result in
secular terms and becomes unreliable at long times. A resummation is necessary
to capture the long time behavior consistently. We will perform a
non-perturbative resummation of the one-loop terms by solving the set of
equations (\ref{umodes} - \ref{onelupeqn}) with the {\em full}
value of $\phi$ in the mode equations.
This then becomes a set of coupled non-linear
integro-differential equations that provide a non-perturbative resummation of
select one-loop terms as can be seen by looking at a perturbative expansion of
the solution to these equations.

The integration of these coupled set of one-loop equations provide the lowest
{\em non-perturbative}self-consistent resummation that is energy
conserving (see below
for energy conservation). This one loop approximation becomes exact
in the large (N,M) limit with N scalar $\sigma$
fields and M fermionic fields when the $\sigma$ field self-interaction is
neglected.

We want to emphasize this point. By incorporating the full value of $\phi$ in
the evolution equation for the mode functions we are incorporating
back-reaction effects self-consistently.  If only the classical evolution of
$\phi$ is used in
the mode equations, we would have parametric resonant amplification since the
effective frequencies for the mode functions are periodic due to the fact that
the classical solution is periodic with constant amplitude. This leads to
particle production that never shuts-off. However particle production leads to
damping in the evolution of $\phi$. Introducing this damped evolution in the
mode equations leads to a behavior rather different from parametric resonance:
as the evolution of $\phi$ is damped, the amplitude becomes smaller
and particle production should diminish and eventually stop.
This was found to be the situation in the case of the scalar field with
self-interaction\cite{reheating}.

Clearly the integration of this set of equations will have to be done
numerically. Because the self-interacting scalar case was already studied
before\cite{reheating}, we will now concentrate on the interaction with the
scalar fields in which thresholds are present. But before we do this we must
understand the renormalization aspects of this system of equations.

The large-$k$ behavior of the mode functions can be understood from a WKB
analysis, the details of which had already been discussed
in\cite{tadpole,reheating}. We find the following divergence structure doing
this
\begin{eqnarray}
&& \int^{\Lambda} \frac{d^3k}{(2\pi)^3} \frac{|U_k(t)|^2}{2W_k}=
\frac{\Lambda}{8\pi^2} - \frac{1}{8\pi^2}
\left(m^2_{\Phi,b}+\frac{\lambda_b}{2}\phi^2(t)\right)\ln\frac{\Lambda}{\kappa}
 + F_1(t,\kappa)\;, \label{umodesdiv}\nonumber \\
&& \int^{\Lambda} \frac{d^3k}{(2\pi)^3} \frac{|V_k(t)|^2}{2w_k}=
\frac{\Lambda}{8\pi^2} - \frac{1}{8\pi^2}
\left(m^2_{\sigma,b}+g_b\phi^2(t)\right)
\ln\frac{\Lambda}{\kappa}
 + F_2 (t,\kappa)\;, \label{vmodesdiv}\nonumber
\end{eqnarray}
where $\Lambda$ is an upper momentum cut-off and $\kappa$ and arbitrary
renormalization scale. The subscript $b$ refers to bare quantities and the
quantities $F_{1,2}(t,\kappa)$ are finite in the limit $\Lambda \rightarrow
\infty$. We can now read the mass and coupling constant renormalizations
\begin{eqnarray}
&& m^2_{\Phi,R}= m^2_{\Phi,b}+ \frac{\Lambda}{8\pi^2}
\left(\frac{\lambda_b}{2} +g _b
\right) -\frac{1}{8\pi^2} \left( \frac{\lambda_b}{2}\; m^2_{\Phi,b}+
g _b\; m^2_{\sigma,b}
\right)\ln\frac{\Lambda}{\kappa}\;, \label{massrenorma}\nonumber \\
&& {\lambda_R} =  {\lambda_b}-\frac{3}{4\pi^2}
\left(  \frac{{\lambda_b}^2}{4}+g ^2_b
\right)\ln\frac{\Lambda}{\kappa} \;. \label{couplingrenorma}\nonumber
\end{eqnarray}
We introduce a further, finite renormalization, by subtracting the functions
$F_{1,2}(t=0,\kappa)$, absorbing this subtraction into a (finite)
renormalization of the mass. After these renormalizations, we finally arrive
at the renormalized set of evolution equations in terms of renormalized
quantities.  We drop the subscript $R$ for renormalized quantities to avoid
cluttering the notation, but with the understanding that all quantities are
renormalized at the scale $\kappa$:
\begin{eqnarray}
&& \ddot{\phi}(t)+ m^2_{\Phi}\, \phi(t)+ \frac{\lambda}{6}\, \phi^3(t)+
\frac{\lambda}{2}\; \phi(t)  \int^{\Lambda} \frac{d^3k}{(2\pi)^3}
\frac{|U_k(t)|^2 -1}{2W_k}
+ g \; \phi(t) \int^{\Lambda} \frac{d^3k}{(2\pi)^3}
\frac{|V_k(t)|^2-1}{2w_k}
\nonumber \\
&& \quad + \frac{1}{8\pi^2}  \left(\frac{{\lambda}^2}{4}+g^2 \right)\phi(t)
\left[\phi^2(t)-\phi^2(0)\right] \ln\frac{\Lambda}{\kappa} =0 \;.
\label{fiequationrenor}
\end{eqnarray}
To this order, we can replace masses and couplings by their renormalized values
in the equations for the mode functions (\ref{umodes}, \ref{vmodes}).
The set of equations (\ref{umodes},\ref{vmodes},\ref{fiequationrenor})
provide a self-consistent, {\em non-perturbative} set of
integro-differential equations with
{\em back-reaction}. This last point is extremely important, a periodic
evolution for the scalar
field in the mode equations (\ref{umodes}, \ref{vmodes}) would lead to
parametric amplification\cite{kofman} and infinite particle
production\cite{yoshimura}.
However the back-reaction of the quantum fluctuations onto the evolution
of the zero mode
of the scalar field is determined completely by these mode equations.
When the amplitude of the
quantum fluctuations grow, this mechanism takes energy of the zero mode,
whose evolution now
becomes damped in amplitude. This damping , a result of the back reaction,
in turn makes the growth of the quantum fluctuations to diminish. Eventually
the particle production mechanism must shut-off as a consequence of this
back reaction.
This is dramatically different from the catastrophic particle production
found by
 Yoshimura\cite{yoshimura} precisely because in that analysis back reaction
was not taken into
account.

We will
now chose the renormalization scale to be $\kappa=m_{\Phi}$ so as to have only
one scale in the problem, which makes the numerical evaluation easier.
Renormalized non-equilibrium equations had also been obtained by
Cooper et. al.\cite{cooper}
within the context of dynamical evolution during the chiral phase transition.
The renormalized equation of motion (\ref{fiequationrenor}) may be written
without reference to the cutoff $\Lambda$ which in the end must be taken to
infinity, however, numerically the $k$-integrals must be calculated with an
upper momentum cutoff. One must ensure that this numerical cutoff, to be
identified with $\Lambda$ in the evolution equation, be much larger than the
masses and amplitudes of the field for the integrals to reach their asymptotics
and the cutoff dependence to dissapear. We have been careful to make exhaustive
checks that the final results were insensitive (to the working accuracy) to the
cutoff which was typically chosen to be $\Lambda \approx 100\, |m_{\Phi}|$.
This implies that the order of magnitude of the error is about
$({{m_{\Phi}}\over {\Lambda}})^2 \simeq 10^{-4}$.

A provision must be made for the initial conditions on the mode functions
solution of (\ref{umodes}, \ref{vmodes}) in the broken symmetry case. In this
case the tree level squared mass is negative $m^2_{\Phi} = -\mu^2$ and for
initial values of the expectation value $\phi^2(0) < 2 \mu^2/ \lambda$ the
initial configuration is below the classical ``spinodal'' and imaginary
frequencies lead to the spinodal instabilities. We are not interested in this
article in studying the time evolution of these instabilities but on the issues
of non-linear relaxation. There are two ways to avoid the complex frequencies
associated with these instabilities in the initial conditions: (i) one can
choose an initial value $\phi^2(0) > 2 \mu^2/ \lambda$ or (ii) one can impose
that the initial frequencies correspond to a positive mass term. This choice
corresponds to preparing an initial gaussian density matrix of the modes with
this given mass, under time evolution this packet spreads in function space and
the time dependence of the width reflects the (quantum and thermal)
fluctuations. In our numerical analysis we chose the later possibility with the
positive mass squared given by the absolute value of the (negative) mass
squared in the Lagrangian.

We introduce now the following dimensionless variables:
\begin{eqnarray}
&& \eta(\tau) =  \sqrt{\frac{\lambda}{6\,|m^2_{\Phi}|}}\phi(t)
\;,\quad\quad\quad
q=\frac{k}{|m_{\Phi}|}\;,\quad\quad\quad
\tau= |m_{\Phi}|\,t \label{dimlessvars1}\;,\quad\quad\quad  \nonumber \\
&& \bar{W}_q= \sqrt{q^2+{|M^2(0)| \over |m^2_{\Phi}|}}\;,\quad\quad\quad
\bar{w}_q= \sqrt{q^2+{m^2(0) \over |m^2_{\Phi}|}}\;. \label{dimlessvars2}
\nonumber
\end{eqnarray}
In terms of which the evolution equation (\ref{fiequationrenor}) becomes
\begin{eqnarray}
&& \ddot{\eta}(\tau)+\eta(\tau)+\eta^3(\tau)+
\frac{\lambda}{8\pi^2}\, \eta(\tau)  \int^{\frac{\Lambda}{|m_{\Phi}|}} q^2{dq}
\,\frac{|U_q(\tau)|^2 -1}{\bar{W}_q}
+\frac{g}{4\pi^2} \; \eta(\tau) \int^{\frac{\Lambda}{|m_{\Phi}|}}q^2 {dq} \,
\frac{|V_q(\tau)|^2-1}{\bar{w}_q}
\nonumber \\
&& \quad\quad\quad\quad +\frac{\lambda}{8\pi^2}
\left({3 \over 2}+{6 g ^2 \over \lambda^2} \right)
\eta(\tau)\left[\eta^2(\tau)-\eta^2(0)\right]  \ln\frac{\Lambda}{|m_{\Phi}|}
=0\;, \label{dimlessfiequationrenor}
\end{eqnarray}
where we chose the renormalization scale $\kappa=|m_{\Phi}|$.

\subsubsection{\bf Particle Production}
As in any time dependent situation, the concept of particle is ambiguous and
has to be specified with respect to some particular state. We choose that state
as the equilibrium ensemble at the initial time $t=0$. The initial condition on
the mode functions for the fluctuations (\ref{umodes}, \ref{vmodes}) naturally
determine the set of positive and negative energy states at this initial
time. At this time the fluctuation operators may be expanded in this basis. The
Fourier components of the fluctuation operators are thus written as
\begin{eqnarray}
&&\chi_{\vec{k}}(0) = \frac{1}{\sqrt{2 W_k}}\left[ a_{\vec{k}}(0)-
a^{\dagger}_{\vec{k}}(0) \right]\;,
\label{creationa}  \nonumber\\
&&\sigma_{\vec{k}}(0) = \frac{1}{\sqrt{2 w_k}}\left[ b_{\vec{k}}(0)-
b^{\dagger}_{\vec{k}}(0) \right]\;. \label{creationb}\nonumber
\end{eqnarray}
The number operators at any time $t$ are
\begin{eqnarray}
&& N_{\chi,k}(t) =
\frac{{\mbox{Tr\,}} a^{\dagger}_{\vec{k}}(0) a_{\vec{k}}(0)
\rho(t)}{{\mbox{Tr\,}}\rho(0)} =
\frac{{\mbox{Tr\,}} a^{\dagger}_{\vec{k}}(t) a_{\vec{k}}(t)
\rho(0)}{{\mbox{Tr\,}}\rho(0)}\;,
\nonumber \\
&& a_{\vec{k}}(t)= U(t)a_{\vec{k}}(0)U^{-1}(t)\;,
\label{nofchi}  \nonumber\\
&& N_{\sigma,k}(t) =
\frac{{\mbox{Tr\,}} b^{\dagger}_{\vec{k}}(0) b_{\vec{k}}(0)
\rho(t)}{{\mbox{Tr\,}}\rho(0)} =
\frac{{\mbox{Tr\,}} b^{\dagger}_{\vec{k}}(t) b_{\vec{k}}(t)
\rho(0)}{{\mbox{Tr\,}}\rho(0)} \;,
\nonumber \\
&&b_{\vec{k}}(t)= U(t)b_{\vec{k}}(0)U^{-1}(t)\;,
\label{nofsigma}\nonumber
\end{eqnarray}
with $U(t)$ the unitary time evolution operator. Following the arguments
presented in reference\cite{reheating} we find that the creation and
annihilation operators at time $t$ are related to those at the initial time
$t=0$ by a Bogoliubov transformation. In terms of the dimensionless variables
introduced above we find
\begin{eqnarray}
&& N_{\chi,q}(\tau)  = \left(2 {\cal{F}}-1\right) N_{\chi,q}(0)
+\left({\cal{F}}-1 \right)\;,
\nonumber \\
&& {\cal{F}}=  \frac{1}{4}|U_q(\tau)|^2\left[1+
\frac{|\dot{U}_q(\tau)|^2}{\bar{W}^2_q|U_q(\tau)|^2}\right]+\frac{1}{2}\;,
\label{nchit}  \nonumber\\
&& N_{\sigma,q}(\tau)  = \left(2 {\cal{G}}-1\right) N_{\sigma,q}(0) +
\left({\cal{G}}-1 \right)\;,
\nonumber \\
&& {\cal{G}}=  \frac{1}{4}|V_q(\tau)|^2\left[1+
\frac{|\dot{V}_q(\tau)|^2}{\bar{w}^2_q|V_q(\tau)|^2}\right]+\frac{1}{2}\;,
\label{nsigmat}\nonumber
\end{eqnarray}
where $N_{\chi,\sigma}(0)$ are the occupation numbers at the initial time $t=0$
and in the case given by the Bose-Einstein distribution functions, derivatives
are with respect to the dimensionless variable $\tau$. Since in this section we
will be working at zero temperature, only the last term will contribute to
particle production. This term is recognized as the induced
contribution. It can be seen from the initial conditions on the wave functions
that the induced contribution vanishes at the initial time.

We will compute numerically the number of particles produced in a correlation
volume
\begin{equation}
{\cal{N}}^b(\tau) = \int d^3 k N_k(t) / |m_{\Phi}|^3 \;.
\nonumber
\end{equation}

\subsubsection{\bf  Energy Conservation:}

To one loop order the expectation value of the Hamiltonian (divided by the
volume)
gives the mean energy density

\begin{eqnarray}
E = &  &\frac{1}{2}\dot{\phi}^2+\frac{1}{2}m^2_{\phi}\phi^2+
\frac{\lambda}{4!} \phi^4+
       \frac{1}{2}\int \frac{d^3k}{(2\pi)^3}\left\{\frac{1}{2W_k(0)} \left[
|\dot{U}_k|^2+W_k^2(t)|U_k|^2 \right]+ \right.  \nonumber \\
       &  & \left.  \frac{1}{2w_k(0)} \left[
|\dot{V}_k|^2+w_k^2(t)|V_k|^2\right]\right\} \label{energy} \\
W_k^2(t)
       &=& k^2+M^2(t) \\
w_k^2(t)
       &=& k^2 + m^2(t)
\end{eqnarray}

Using the one-loop equations for the zero mode and the mode functions it is
straightforward to find that the above energy density is conserved in time.
Thus this
(resummed) one-loop set of self-consistent equations is energy conserving.
This is
an important consistency check of the set of equations that guarantees
that the numerical results are trustworthy.

\subsection{Inflaton Coupled to Fermions Only}
The fermionic Green's functions are constructed from the solutions to the
homogeneous Dirac equation in the presence of the background field.
The main ingredients and treatment is similar to that of fermions in
presence of an electric
field studied by Kluger et. al.\cite{kluger}.
Writing
the independent solutions as
\begin{eqnarray}
&&{\cal{U}}^{(1,2)}(\vec{x},t)= e^{i\vec{k}\cdot\vec{x}}U^{(1,2)}_k(t)\;,
\nonumber \\
&&{\cal{V}}^{(1,2)}(\vec{x},t)= e^{-i\vec{k}\cdot\vec{x}}V^{(1,2)}_k(t)\;,
\nonumber
\end{eqnarray}
the mode functions obey
\begin{eqnarray}
&&\left[i\gamma_0\frac{d}{dt}-\vec{\gamma}\cdot\vec{k}-m_{\psi}(t)\right]
U^{(1,2)}_k(t)=0\;, \label{uspinor}  \nonumber\\
&&\left[i\gamma_0\frac{d}{dt}+\vec{\gamma}\cdot\vec{k}-m_{\psi}(t)\right]
V^{(1,2)}_k(t)=0\;. \label{vspinor}\nonumber
\end{eqnarray}
It is convenient to write the spinors as
\begin{eqnarray}
&&U^{(1,2)}_k(t) =
\left[i\gamma_0\frac{d}{dt}-
\vec{\gamma}\cdot\vec{k}+m_{\psi}(t)\right]f_k(t)u^{(1,2)}\;,
\label{ffunct}  \nonumber\\
&&V^{(1,2)}_k(t) =
\left[i\gamma_0\frac{d}{dt}+
\vec{\gamma}\cdot\vec{k}+m_{\psi}(t)\right]g_k(t)v^{(1,2)}\;,
\label{gfunct}\nonumber
\end{eqnarray}
with $u^{(1,2)}$, $v^{(1,2)}$ the spinor eigenstates of $\gamma_0$ with
eigenvalues $+1$, $-1$ respectively. The functions $f_k(t)$, $g_k(t)$ obey
the second order equations
\begin{eqnarray}
\left[\frac{d^2}{dt^2}+\vec{k}^2+m^2_{\psi}(t)-i\dot{m}_{\psi}(t)\right]
f_k(t)=0\;, \label{fequation} \nonumber\\
\left[\frac{d^2}{dt^2}+\vec{k}^2+m^2_{\psi}(t)+i\dot{m}_{\psi}(t)\right]
g_k(t)=0\;. \label{gequation}\nonumber
\end{eqnarray}
We now need to append initial conditions. We will consider the situation in
which the system was in equilibrium at time $t \leq 0$ with the expectation
value of scalar field being $\phi(0)$ and $\dot{\phi}(0)=0$.
 Thus the fermion mass is constant and
given by $m_{\psi}(0)$. We can now impose the condition that the modes
$f_k(t)$, $g_k(t)$ describe positive and negative frequency solutions for
$t\leq0$ and, normalizing the spinor solutions to the Dirac equation to unity,
we impose the following initial conditions
\begin{eqnarray}
&&f_k(t<0)= \frac{e^{-ik_0t}}{\sqrt{2k_0 (k_0+m_{\psi}(0) ) }} \;,
\label{bcfoft}  \nonumber\\
&&g_k(t<0)= \frac{e^{ ik_0t}}{\sqrt{2k_0 (k_0+m_{\psi}(0) ) }} \;,
\label{bcgoft} \nonumber\\
&& k_0= \sqrt{{\vec{k}}^2+m^2_{\psi}(0)}\;. \label{k0} \nonumber
\end{eqnarray}
Equations (\ref{gequation}) with these boundary conditions imply that
\begin{equation}
g_k(t)=f^*_k(t)\;. 
\nonumber
\end{equation}
The necessary ingredients for the zero temperature fermionic Green's
functions are the following
\begin{eqnarray}
&& S^>_k(t,t')= -i \sum_{\alpha=1,2}U^{\alpha}_{\vec{k}}(t)
 \bar{U}^{\alpha}_{\vec{k}}(t')\;,
 \label{sgreat}  \nonumber\\
&& S^<_k(t,t')= i \sum_{\alpha=1,2}V^{\alpha}_{-\vec{k}}(t)
 \bar{V}^{\alpha}_{-\vec{k}}(t') \;. \label{sles}\nonumber
\end{eqnarray}
In the standard Dirac representation for the $\gamma$ matrices, we find
\begin{eqnarray}
&&S^>_k(t,t')= -i f_k(t)f^*_k(t')\left[{\cal{W}}_k(t)\gamma_0-
\vec{\gamma}\cdot
\vec{k}+m_{\psi}(t)\right]\left(\frac{1+\gamma_0}{2}\right)
\left[{\cal{W}}^*_k(t')\gamma_0-\vec{\gamma}\cdot
\vec{k}+m_{\psi}(t')\right]\;, \label{sgreatoft} \nonumber \\
&&S^<_k(t,t')= -i f^*_k(t)f_k(t')\left[{\cal{W}}^*_k(t)\gamma_0-
\vec{\gamma}\cdot
\vec{k}-m_{\psi}(t)\right]\left(\frac{1-\gamma_0}{2}\right)
\left[{\cal{W}}_k(t')\gamma_0-\vec{\gamma}\cdot
\vec{k}-m_{\psi}(t')\right]\;, \label{slessoft}  \nonumber\\
&&\quad {\cal{W}}_k(t)= i \frac{\dot{f}_k(t)}{f_k(t)}\;.
\label{timefreq}\nonumber
\end{eqnarray}
With these ingredients the non-equilibrium fermionic Green's functions can be
constructed as in section II. It is an important check that the equality
${\mbox{Tr\,}} S^{++}_k(t,t)= {\mbox{Tr\,}} S^{--}_k(t,t)=
{\mbox{Tr\,}} S^{>}_k(t,t)= {\mbox{Tr\,}} S^{<}_k(t,t)$ is
fulfilled where the trace is over Dirac indices. It is also an important
property that
\begin{equation}
\left(-i\dot{f}^*_k(t)+m_{\psi}(t) f^*_k(t)\right)
\left(i\dot{f}_k(t)+m_{\psi}(t)f_k(t)\right)+k^2f^*_k(t)f_k(t)=1\;.
\nonumber
\end{equation}
This is a consequence of the conservation of probability and may be checked
explicitly.

The evolution equation becomes
\begin{eqnarray}
&&\ddot{\phi}(t)+ m^2_{\Phi}\, \phi(t)+ \frac{\lambda}{6}\, \phi^3(t)+
\frac{\lambda}{2}\, \phi(t) <\chi^2(\vec{x},t)> -y \,
{\mbox{Tr\,}} S^<(\vec{x},t;\vec{x},t)=0\;,
\label{onelupeqnfer}  \nonumber\\
&&{\mbox{Tr\,}} S^{<}(\vec{x},t;\vec{x},t) = 2 \int \frac{d^3k}{(2\pi)^3}
 \left[1-2k^2|f_k(t)|^2 \right]\;.
\label{trace}
\end{eqnarray}

The expectation value of the Hamiltonian can again be computed in
this (resummed) one-loop
approximation. Using the equations of motion for the expectation value
of the scalar field
and those of the mode functions of the fermions, it can be shown to be
conserved similarly to
the bosonic case.

The integral in (\ref{trace}) is divergent. The divergence structure may be
understood in two different ways: carrying out a WKB expansion of the modes,
just as was done in the bosonic case (this is a tedious and lengthy exercise),
or alternatively calculating the closed loop in the presence of the background
field. We carried out both methods using an upper momentum cutoff in the
integrals and found
\begin{equation}
2y \int \frac{d^3k}{(2\pi)^3}
 \left[1-2\, k^2|f_k(t)|^2 \right] = -\frac{ y}{2\pi^2}\; \Lambda^2 \;
m_{\psi}(t) +\frac{y}{4\pi^2}[\ddot{m}_{\psi}(t)+2\,  m^3_{\psi}(t)]
\ln\frac{\Lambda}{\kappa} +\mbox{ finite }\;, \label{diverfer}
\end{equation}
where $\kappa$ is an arbitrary renormalization scale which will be chosen again
to be $\kappa=m_{\Phi}$ for numerical convenience. We will concentrate on the
``chiral limit'' in which the {\it bare} mass term for the fermions vanishes
and $m_{\psi}(t)=y\, \phi(t)$. In this limit the discrete symmetry $\phi
\rightarrow -\phi$ is explicit. The divergent terms in (\ref{diverfer}) are
then identified as: mass renormalization (of $\phi$) (first term), wave
function renormalization (second term) and coupling constant renormalization
(third term).  From now on, all quantities will be renormalized, and we drop
the subscript $R$ to avoid cluttering.  For simplicity in the numerical
analysis and
in order to isolate the fluctuations from the fermionic contribution from those
of the scalar sector, we now neglect the contribution from the scalar loop.
The scalar fluctuations had already been analyzed in the previous section and
in\cite{reheating}.  The final renormalized equation of motion is now
\begin{eqnarray}
&&\ddot{\phi}(t)+ m^2_{\Phi}\; \phi(t)+\frac{\lambda}{6}\phi^3(t) -2y
 \int \frac{d^3k}{(2\pi)^3}
 \left[1-2k^2|f_k(t)|^2 \right]-\left\{ -\frac{ y^2}{2\pi^2}\Lambda^2
\; \phi(t)\nonumber \right. \\
&& \left. \quad\quad\quad+\frac{y^2}{4\pi^2}\left[ \ddot{\phi}(t)+
2 y^2 \; \phi^3(t)\right]
\ln\frac{\Lambda}{m_{\Phi}}
 \right\}=0 \label{renoonelupeqnfer}\nonumber\;.
\end{eqnarray}
With the purpose of a numerical analysis of this equations, it proves
convenient to introduce the following dimensionless of variables
\begin{eqnarray}
  && \eta= y\phi/m_{\Phi}\;,\quad \tau= m_{\Phi}t\;,\quad q=k/ m_{\Phi}
\;,\quad g'=\lambda_R/6y^2 \;,\quad g=y^2/\pi^2\;, \nonumber \\
  && u_{q} (\tau) =f_{k} (t) \sqrt{\omega^{0}_{k} (\omega^{0}_{k}+
y^2 \phi^2(0))}\;, \nonumber
\end{eqnarray}
in terms of which the equations of motion  are
\begin{eqnarray}
 && \frac{d^2 \eta}{d \tau^2} +\eta +g' \eta^3 -g \Sigma (\tau) =0\;,
\label{fermeq1} \nonumber\\
&& \left[ \frac{d^2}{d \tau^2} +q^2 +\eta^2 -i \dot{\eta} \right] u_{q}
 (\tau)=0\;,\quad\quad u_q(0) = \frac{1}{\sqrt{2}}\;,\quad\quad\dot{u}_q(0)=
-i{\sqrt{q^2+\eta^2(0)}\over \sqrt{2}} \label{fermeq2}\;,  \nonumber \\
&& \Sigma(\tau) = \int^{\Lambda/m_{\Phi}}_{0}  q^2 \, dq
\left[ 1- {q^2 |u_{q} (\tau)|^2
\over \sqrt{q^2+\eta^2 (0)} ( \sqrt{q^2+\eta^2 (0)} +\eta(0) )}
 \right]\nonumber \\
 &&\quad\quad\quad
 -{1\over2} \left( {\Lambda \over m_{R,\Phi}} \right)^2 \eta +
{1 \over 4}\left( \ddot{\eta} + 2 \, \eta^3 \right) \ln{\Lambda \over
m_{R,\Phi}}\;.
\label{finaleqferm}
\end{eqnarray}
Although the equations are finite when the ultraviolet cutoff is taken to
infinity and can be written without the introduction of the cutoff by
subtracting the integral with a {\it lower} limit cutoff given by the
renormalization scale, numerically these integrals will have to be done by
introducing an upper momentum cutoff anyways. Therefore we keep this UV cutoff
in the equations.

\subsubsection{\bf Particle Production}
The study of particle production is similar to the scalar case, with the
only complication being the spinorial structure of the fermionic fields.
At the initial time $t=0$ the quantized fermion operator is
\begin{equation}
 \psi ( \vec{x}, 0) =\int {d^3 k \over (2\pi)^3} \sum_{\alpha =1,2}
\left[
b^{(\alpha)}_{k}(0) U^{(\alpha)}_{k} (0) + d^{(\alpha)\dagger}_{-k}(0)
 V^{(\alpha)}_{-k} (0) \right] e^{i \vec{k} \cdot \vec{x}}\;,\nonumber
\end{equation}
in terms of the mode functions determined above. The number of fermions
and antifermions are {\it defined} as
\begin{eqnarray}
&& \langle N^f(t) \rangle = \int \frac{d^3k}{(2\pi)^3} \sum_{\alpha}
\frac{{\mbox{Tr\,}} \left[\rho(t) b^{ (\alpha) \dagger}_{k} (0)
 b^{(\alpha)}_{k} (0)\right]}{{\mbox{Tr\,}} \rho(0)}
= \int \frac{d^3k}{(2\pi)^3} \sum_{\alpha}
\frac{{\mbox{Tr\,}} \left[\rho(0) b^{ (\alpha) \dagger}_{k} (t)
b^{(\alpha)}_{k} (t)\right]}{{\mbox{Tr\,}} \rho(0)} \;,
\label{fermionnumb} \nonumber \\
&& \langle {N}^{\bar{f}} (t) \rangle =\int {d^3k \over
(2\pi)^3} \sum_{\alpha}
\frac{{\mbox{Tr\,}} \left[ \rho(t) d^{ (\alpha) \dagger}_{k} (0)
d^{(\alpha)}_{k} (0)\right]}{{\mbox{Tr\,}} \rho(0)}
= \int \frac{d^3k}{(2\pi)^3} \sum_{\alpha}
\frac{{\mbox{Tr\,}} \left[\rho(0) d^{ (\alpha) \dagger}_{k} (t)
 d^{(\alpha)}_{k} (t)\right]}{{\mbox{Tr\,}} \rho(0)} \;.
\label{antifernum}\nonumber
\end{eqnarray}
The time dependent coefficients are obtained by projecting the time dependent
spinor solutions onto the positive and negative energy solutions at $t=0$, and
are related to the coefficients at $t=0$ via a Bogoliubov transformation
\begin{eqnarray}
&&  b^{(\alpha)}_{k} (t)= \sum_{\beta} \left[ {\cal{F}}^{(\alpha)}_{(\beta)k,+}
  (t)  b^{(\beta)}_{k} +  {\cal{F}}^{(\alpha)}_{(\beta)k,-} (t)
d^{ (\beta) \dagger}_{-k}
  \right]\;,  \nonumber\\
&&  d^{ (\alpha)\dagger}_{-k} (t)= \sum_{\beta}
\left[ {\cal{H}}^{(\alpha)}_{(\beta)k,+}
  (t)  b^{(\beta)}_{k} +  {\cal{H}}^{(\alpha)}_{(\beta)k,-}(t)
 d^{(\beta)\dagger}_{-k}  \right]\;.  \nonumber
\end{eqnarray}
The identities
\begin{eqnarray}
  &&\sum_{\beta} \mid {\cal{F}}^{(\alpha)}_{(\beta)k,+}
  (t) \mid^2 + \mid {\cal{F}}^{(\alpha)}_{(\beta)k,-}
  (t) \mid^2 =1\nonumber\;, \\
   &&\sum_{ \beta} \mid {\cal{H}}^{(\alpha)}_{(\beta)k,+}
  (t) \mid^2 + \mid {\cal{H}}^{(\alpha)}_{(\beta)k,-}
  (t) \mid^2 =1 \nonumber
\end{eqnarray}
ensure that the transformations preserve the anticommutation relations as they
must. It is also a matter of algebra using the equations of motion for the
mode functions to prove that the number of fermions minus antifermions is
conserved for each $ \vec{k}$ mode. In what follows, we only consider the
number of fermions produced since fermions and antifermions are created in
pairs.

With some algebra, the Bogoliubov coefficients can be expressed as the function
of $f_k(t)$ and $f^{*}_k (t)$.  In terms of the dimensionless variables defined
above we obtain the number of fermions within a correlation volume
${\cal{N}}^{f} (\tau) = N^{f}(\tau)/{|m_{\Phi}|^3}$
\begin{eqnarray}
  {\cal{N}}^{f}(\tau) &&= {1 \over 2\pi^2} \int q^2\, dq {\cal{N}}^{f}_{q}
(\tau) \nonumber \\
          &&= {1 \over 4 \pi^2} \int dq \left[ \frac{q^2}{ \sqrt{q^2+\eta^2(0)}
           \left( \sqrt{q^2+\eta^2(0)}+\eta(0) \right) } \right]^2 \nonumber \\
   &&  ~~~ \times \left[ -i \frac{\partial}{\partial \tau} +\eta(\tau) -
 \sqrt{q^2+\eta^2(0)}-\eta(0)  \right] u^{*}_{q} (\tau) \nonumber \\
    &&  ~~~ \times \left[ i \frac{\partial}{ \partial \tau} +\eta(\tau) -
 \sqrt{q^2+\eta^2(0)}-\eta(0)  \right] u_{q} (\tau) \label{ferminumb}\;.
\end{eqnarray}

\section{\bf Numerical Analysis}
The numerical analysis was carried out with a fourth order Runge-Kutta method
for the differential equations and a 5-point Bode rule integrator for the
$k$-integrals. The typical step size in time was $1-2.10^{-3}$, and the typical
step size in $k$ was $10^{-3}$. The cutoff $\Lambda/|m_{\Phi}| $ was varied
between 75 and 200; we found no sensitivity to the cutoff in the range of
parameters that we tested (see below). The code was very stable within the
range of parameters tested.

\subsection{\bf Inflaton Coupled to Scalars: Unbroken Symmetry Case}
Since the contribution of the one-loop quantum fluctuations of the inflaton
$\Phi$ have already been studied previously\cite{reheating} and we want to
study the contribution of the lighter field $\sigma$, we will neglect the
contribution from the self-interaction in the evolution equation. Thus we study
the following equations obtained from equations
(\ref{vmodes}, \ref{dimlessfiequationrenor}) in terms of
dimensionless variables
\begin{eqnarray}
&& \ddot{\eta}(\tau)+\eta(\tau)+\eta^3(\tau)+
 \frac{g}{4\pi^2} \; \eta(\tau) \int^{\frac{\Lambda}{|m_{\Phi}|}}q^2 {dq} \;
\frac{|V_q(\tau)|^2-1}{\bar{w}_q}
 \nonumber \\
&& \quad\quad\quad +\frac{\lambda}{8\pi^2}
 \;{6 g ^2 \over \lambda^2} \;  \eta(\tau)\left[\eta^2(\tau)-\eta^2(0)\right]
 \ln\frac{\Lambda}{|m_{\Phi}|}=0\;, \label{reducedeq2}  \nonumber\\
&&\left[\frac{d^2}{d\tau^2}+\vec{q}^2+\frac{m^2_{\sigma}}{|m_{\Phi}|^2}
+\frac{6g}{\lambda}\eta(\tau) \right] V_q(\tau)=0\;, \nonumber \\
&& V_k(0)=1\;, ~~~ \dot{V}_q(0)= -i \, \bar{w}_q= -i \, \sqrt{{\vec{q}}^2+
\frac{m^2_{\sigma}}{|m_{\Phi}|^2}
+\frac{6g}{\lambda}\eta^2(0) }\;. \nonumber
\end{eqnarray}
Notice that the factor (3/2) multiplying the logarithm in
eq. (\ref{dimlessfiequationrenor}) and missing from (\ref{reducedeq2}) arises
from the renormalization of the $\Phi$-scalar loop that is not taken into
account in (\ref{reducedeq2}).

Figures (1.a-c) show the unbroken symmetry case with the quantum fluctuations
from the scalar loop of the $\sigma$ particles for the values $y=0;~~ \lambda
/8\pi^2=0.2;~~g=\lambda; ~~ m_{\sigma}=0.2\,m_{\Phi};~~
\eta(0)=1.0;~~\dot{\eta}(0)=0$.  Figure (1.a) shows $\eta(\tau)$ vs $\tau$,
figure (1.b) shows ${\cal{N}}_{\sigma}(\tau)$ vs $\tau$ and figure (1.c) shows
${\cal{N}}_{q,\sigma}$ vs $q$ for $\tau=120$, similar graphs were obtained with
snapshots at different (earlier) times.

Figure (1.a) shows a very rapid, non-exponential damping within few
oscillations of the expectation value and a saturation effect when the
amplitude of the oscillation is rather small (about 0.1 in this case), the
amplitude remains almost constant at the latest times tested. Figure (1.a) and
figure (1.b) clearly show that the time scale for dissipation (from figure
(1.a) is that for which the particle production mechanism is more efficient
(figure (1.b)). Notice that the total number of particles produced rises on the
same time scale as that of damping in figure (1.a) and eventually when the
expectation value oscillates with (almost) constant amplitude the average
number of particles produced remains constant. These figures clearly show that
damping is a consequence of particle production. At times larger than about 40
$m_{\Phi}^{-1}$ (for the initial values and couplings chosen) there is no
appreciable damping. The amplitude is rather small and particle production has
practically shut off. If we had used the {\it classical} evolution of the
expectation value in the mode equations, particle production would not shut off
(parametric resonant amplification), and thus we clearly see the dramatic
effects of the inclusion of the back reaction.

In this unbroken symmetry case linear relaxation predicts a slow
$t^{-3}$ power law
decay to an asymptotic finite amplitude because one particle decay is
kinematically forbidden: the self energy contribution is a two-loop effect with
one $\Phi$ and two $\sigma$  particle cut and kinematically there is a
one-particle
pole below the three particle threshold. A slow power law linear relaxation
asymptotically cannot be ruled out numerically because we have not continued
the integration for longer times but clearly asymptotically the numerical
result is compatible with linear relaxation. Figure (1.c) shows the
distribution of particles created at the latest time $\tau =120$ as a function
of wave vector. Similar graphs were obtained with snapshots at different
earlier times. The distribution is clearly non-thermal and skewed towards small
momentum (in units of $m_{\Phi}$).  These figures are qualitatively similar to
those obtained in the self-interacting case in\cite{reheating}. The asymptotic
behavior is that of undamped oscillations of small amplitude. This is
compatible with the result from the linear relaxation analysis because there is
a one-particle pole below the three particle threshold resulting in undamped
oscillations at large times. Linear relaxation predicts qualitatively the same
results for the self-interacting scalar case and this case in the unbroken
phase. The numerical results are consistent with this prediction.

However, for large amplitudes non-linear relaxation via particle production is
very effective and dramatically different from linear relaxation.

\subsection{\bf Inflaton Coupled to Scalars: Broken Symmetry Case}
In this case, we take $m^2_{\Phi}= - |m^2_{\Phi}|$.

As in the unbroken symmetry case, we only study the effect of the lighter
$\sigma$ fluctuations. In the broken symmetry case, linear relaxation predicts
an open decay channel for the inflaton, resulting in exponential relaxation for
a long time and eventually relaxation with a power law. Therefore we should not
expect a constant amplitude asymptotically but an amplitude that eventually
should relax to zero. In this case the renormalized equations for evolution are
\begin{eqnarray}
&& \ddot{\eta}(\tau)-\eta(\tau)+\eta^3(\tau)+
 \frac{g}{4\pi^2} \; \eta(\tau) \int^{\frac{\Lambda}{|m_{\Phi}|}}q^2 {dq}
\; \frac{|V_q(\tau)|^2-1}{\bar{w}_q}
 \nonumber \\
&& \quad\quad\quad +\frac{\lambda}{8\pi^2}
 \; {6 g ^2 \over \lambda^2} \; \eta(\tau)\left[\eta^2(\tau)-\eta^2(0)\right]
 \ln\frac{\Lambda}{|m_{\Phi}|}=0\;, \label{reducedeq}  \nonumber\\
&&\left[\frac{d^2}{d\tau^2}+\vec{q}^2+\frac{m^2_{\sigma}}{|m_{\Phi}|^2}
+\frac{6g}{\lambda}\eta(\tau) \right] V_q(\tau)=0\;,\nonumber  \\
&& V_k(0)=1\;, ~~~ \dot{V}_q(0)= -i \, \bar{w}_q= -i \sqrt{\vec{q}^2+
\frac{m^2_{\sigma}}{|m_{\Phi}|^2}+\frac{6g}{\lambda}\eta^2(0) } \;. \nonumber
\end{eqnarray}
Figure (2.a-c) show $\eta(\tau)$ vs $\tau$, ${\cal{N}}_{\sigma}(\tau)$ vs
$\tau$ and ${\cal{N}}_{q,\sigma}(\tau=200)$ vs $q$ respectively, for
$\lambda / 8\pi^2 =
0.2;~~~ g / \lambda = 0.05;~~~ m_{\sigma}= 0.2\, |m_{\Phi}|; ~~
\eta(0)=0.6;~~~\dot{\eta}(0)=0$. Notice that the mass for the linearized
perturbations of the $\Phi$ field at the broken symmetry ground state is
$\sqrt{2}\,|m_{\Phi}| > 2 m_{\sigma}$. Therefore, for the values used in the
numerical analysis, the two-particle decay channel is open for linear
relaxation. For these values of the parameters, linear relaxation predicts
exponential decay with a time scale $\tau_{rel} \approx 300$ (in the units
used). Figure (2.a) shows very rapid non-exponential damping on time scales
about {\em six times shorter} than that predicted by linear relaxation. The
expectation value reaches very rapidly a small amplitude regime, once this
happens its amplitude relaxes very slowly. Within our computing time
limitations we could not confirm that there is exponential relaxation in the
small amplitude regime (for $\tau > 100$) but clearly there is a striking
difference with the unbroken symmetry case. The influence of open channels is
evident, however in the non-linear regime relaxation is clearly {\em not}
exponential but
extremely fast. Although we cannot confirm the exponential (or power law)
relaxation numerically in the small amplitude regime,
 the amplitude at long times seems to relax to the
expected value, shifted slightly from the minimum of the tree level potential
at $\eta = 1$. This is as expected from the fact that there are quantum
fluctuations. Figure (2.b) shows that particle production occurs during the
time scale for which dissipation is most effective, giving direct proof that
dissipation is a consequence of particle production. Asymptotically, when the
amplitude of the expectation value is small, particle production shuts off. We
point out again that this is a consequence of the back-reaction in the
evolution equations. Without this back-reaction, as argued above, particle
production would continue without indefinitely. Figure (2.c) shows that the
distribution of produced particles is very far from thermal and concentrated at
low momentum modes $k \leq |m_{\Phi}|$. This distribution is qualitatively
similar to that in the unbroken symmetry case, and points out that the excited
state obtained asymptotically is far from thermal.

\subsection{\bf Inflaton Coupled to Fermions Only: Unbroken Symmetry Case}
Here we treat the case $y \neq 0;~~g=0$.

The renormalized evolution and particle production equations in this case are
given by eq. (\ref{finaleqferm}) and
eq. (\ref{ferminumb}), respectively, in terms of dimensionless
quantities. Figures (3.a-c) show $\eta(\tau)$ vs. $\tau$, ${\cal{N}}^f(\tau)$
vs $\tau$ and ${\cal{N}}^f_q(\tau=200)$ respectively for the values of the
parameters $m_{\psi}=0;~~y^2 / \pi^2 = 0.5;~~ \lambda / 6y^2 = 1.0;~~
\eta(0)=0.6;~~\dot{\eta}(0)=0$. One observes from these figures that after a
rather brief period of initial damping of just a few oscillations of the scalar
field, the dissipative mechanism shuts-off. Figure (3.b) shows that during this
time scale fermion-antifermion pairs are being produced but then the number of
produced particles saturates and oscillates with a small amplitude. Figure
(3.c) shows that the distribution of fermions produced is peaked at very low
momentum ($k \leq m_{\Phi}$) and with a maximum value of 2, which is the total
number of degrees of freedom per $k$-wave vector. These numerical results
expose the physics of Pauli blocking very clearly; the available low momentum
modes are occupied and no more fermion-antifermion pairs can be produced. Pauli
blocking shuts off particle production and dissipation very early on. We have
obtained snapshots of the particle number as a function of momentum for
different times, and they all present the same picture.

This result is markedly different from the prediction of linear relaxation. At
zero temperature, eq. (\ref{widthfer}) for linear relaxation predicts
exponential damping with a (dimensionless) time scale $\tau_{rel} \approx 10.6$
for the values of the parameters chosen above. The difference between the
non-linear evolution and that predicted by linear relaxation is explained by
the Pauli blocking phenomenon.  Very early in the evolution, the low momentum
available fermionic states were filled with produced fermions.  Once these
states have filled, damping and particle production shuts off. This Pauli
blocking effect is explicit in eq. (\ref{widthfer}) but there it appears
from finite temperature effects. In the non-linear relaxation case, we began at
zero temperature, but an excited state quickly ensues because of particle
production.

This analysis reveals that for large amplitudes of the scalar field, fermions
will be rather ineffective in dissipation and damping because of Pauli blocking
{\em even at zero temperature}. The time scales for dissipation obtained from
the fermion self-energies, which apply to the case of linear relaxation are
completely unrelated to the time scales for non-linear dissipative processes,
even asymptotically, because the fermionic states are Pauli blocked.

\subsection{\bf  Thermalization: the argument}

The self-consistent resummed one-loop approximation used in our analysis
does not
incorporate collisional processes. These will appear in the next
(two loop order). The
distribution of the particles produced is clearly very far from
thermal, with typically very
many particles in the low momentum modes $k \leq m_{\phi}$.
Although a full study of
thermalization will necessarily involve the two loop contribution
(and consequently memory
effects that will ultimately result in an extensive numerical effort),
we can argue on the
basis of an effective Boltzmann description. We can take the distribution
of particles obtained
from the numerical analysis at long times (times longer than the
non-linear relaxation scale) as  an input in a Boltzmann equation.
The integration of this Boltzmann equation will determine the
time scales for collisional thermalization. Because the couplings will
be typically very
small and the initial occupations are rather far from thermal equilibrium
these collisional time
scales will be very long (as compared to the non-linear relaxation scales).
Collisional relaxation
does not change the number of particles and is energy conserving
(in flat Minkowsky space),
just re-distributes the occupation. Therefore a thermal
equilibrium state will result with the total number of particles
given by the number of produced
particles obtained during the fast period of induced amplification
and the final reheating temperature will be determined by this
number of particles. Including the collisional relaxation
via the Boltzmann equation during the period of induced amplification will
not modify the time
scales substantially. The reason is that initially the occupation
numbers are rather small and
Boltzmann processes are suppressed. Since the rate of particle production
depends on the couplings {\em and} the initial amplitudes, for large values
of the amplitudes the rate of particle
production will be much larger than the rate of equilibration via Boltzmann
processes (second
order in the couplings and independent of the initial amplitude). Thus
collisional (Boltzmann) processes will become relevant for  thermalization
at times
much longer (for weak coupling) than the times during which most of the
particle production
takes place.
If the rate of thermalization obtained from this Boltzmann equation is
much larger than the
expansion rate of the universe, then the reheating temperature can just
be estimated from the
total number of particles produced during the period of induced
amplification. However if this
collisional thermalization rate is much smaller than the expansion rate
there will be substantial redshifting of the temperature and a full
numerical analysis will have to be carried out
to obtain a reliable estimate of the reheating temperature.

\subsection{\bf Estimate of the Reheating Temperature}

With the arguments on thermalization given above in mind, and {\em assuming}
that the
thermalization rate is much larger than the expansion rate (no red-shifting
of the temperature),
we can use the numerical analysis presented above to obtain an estimate of
the reheating
temperature.

The important conclusion of the numerical integration is that the density of
particles
produced during the period of induced amplification is
\begin{equation}
{\cal{N}}= c \; m_{a}^3
\end{equation}
where $ m_{a} $ is a mass in between the inflaton and the $\sigma$ field
masses.
$c$ a constant of ${\cal{O}}(1-10)$. This constant depends on the
{\em product}
 of the coupling constants and the {\em initial amplitudes} as can be
seen from the
mode equations. It also depends approximately logarithmically on the
coupling constants
because of the time scales for the shutting-off of the induced
amplification mechanism.
This can be seen as follows: from the expression for the number
of (bosonic) particles
produced we find the {\em rate} of particle production (for example
for the $\sigma$
field)
\begin{equation}
\dot{N}_k(t) = \frac{g\phi^2(0)}{w^2_k(0)}
\left[1-\frac{\phi^2(t)}{\phi^2(0)}\right]
\frac{d |V_k(t)|^2}{dt} \label{rateparprod}
\end{equation}
This expression clearly shows that the {\em rate} of particle
production depends on the
produc of  the coupling and the initial amplitude (squared for the
interaction considered),
since the mode functions also depend on this combination. The total
number of particles
is given by this rate integrated up to the time at which the amplification
mechanism is shut-off
by the back reaction. Un upper bound to this time is estimated as follows:
in the early stages, the mode functions increase almost
exponentially (because of the resonances)  this growth is modified by the
damping of the
amplitude of the zero mode. Since we find that the back reaction shuts-off
the particle
production when the one loop term becomes of the order of the tree-level
term, this gives a
time scale that is logarithmic in the coupling. Thus
even for very small couplings but very large initial amplitude of the
inflaton zero mode
(and this is the situation in cosmological scenarios) the constant $c$
will be of
${\cal{O}}(1-10)$ and can be enhanced by giving the inflaton a larger
initial amplitude.
Under the assumption of thermalization on time scales shorter than the
expansion scale
and that the mass of scalars and fermions are much smaller than the
inflaton mass an
estimate of the reheating temperature is obtained as
\begin{equation}
T_{reh} \approx {\cal{N}}^{\frac{1}{3}}\approx m_{a}.
\end{equation}

\section{Conclusions and Implications}
The reheating  and thermalization processes in inflationary universe
models occur very
far from equilibrium and must be studied in their full complexity,
eventually numerically.
The methods developed within real-time non-equilibrium field theory
allow a consistent
treatment that we used to  obtain  the equations of evolution for the
expectation
value of the inflaton field. These equations are non-perturbative and
take into account the
 back reaction effect of
both the self couplings of the inflaton, as well as the couplings to lighter
scalars and fermions.

We have examined these evolution equations, both in the linear regime, in which
the amplitude of the field is small and in the opposite extreme, the non-linear
regime, which is intrinsically non-perturbative.

In the linear regime, we have seen that if the inflaton mass is above the two
particle threshold, there is some damping behavior for a period of time, and
indeed, this damping is governed by the total width of the inflaton. However,
at late times the behavior is dominated by {\it power law} decay rather than
exponential damping. Furthermore, we have noted the important distinction
between the time scale for the relaxation of the expectation value of the
inflaton as opposed to thermalization of the produced particles. It is clear
that as treated in this work, relaxation can occur well before thermalization,
i.e. interactions that drive the particle distributions towards a thermal one,
has a chance to become relevant.

The more interesting case, and the one which would most likely be relevant to
the reheating problem in models such as chaotic inflation\cite{chaotic}, is
that of when the inflaton is in the non-linear regime. Unlike the
elementary approach
to reheating, in which the ``decay'' of the coherent inflaton oscillations
occured a time $t_{\rm decay}\sim \Gamma^{-1}$, where $\Gamma$ is the inflaton
decay width, after the end of the slow-roll regime, particle production in the
non-linear regime begins at very early times and then shuts off. Indeed, by the
time the inflaton expectation value reaches its asymptotic state of
oscillations around its minimum, the period of particle production is
essentially over. This will be the stage of thermalization via collisions,
however we find that
the spectrum of produced particles is extremely non-thermal,
and collisional relaxation towards
an equilibrium thermal state may take a long time.

An inflationary model that cannot reheat the universe to at least
nucleosynthesis temperatures is wrong. Thus it is clearly important to
accurately compute the reheating temperature in inflationary
models. What our analysis here
shows is that this computation is of necessity much more complicated than has
previously been thought but perhaps more importantly that there are
different time scales.
Although so far we have studied the situation in flat Minkowski space,
recent investigations
by Kaiser\cite{kaiser} reveal that these non-linear processes will
still be very effective in
inflationary cosmologies.

The first stage which occurs rather fast is that of particle production
and damping of
oscillations as a result of induced amplification. The second stage,
that begins typically when
the amplitude of the inflaton is rather small and it oscillates with
almost constant amplitude at
the bottom of the potential is that of one-particle decay and of
thermalization via collisional relaxation.

The thermalization time will have to be
studied setting up a (quantum) Boltzmann equation using the distributions
of produced particles
at the end of the induced amplification stage as input for the
Boltzmann evolution.  Since these
initial distributions are very far from equilibrium, it may take many
collisions to relax to a
thermal equilibrium state. If the couplings are very small
(and this will clearly depend on the
models) the thermalization time may be very large.

Following the particle numbers to the point at which they become
thermal would then allow us to pick off the final reheating temperature.
In the weak coupling regime, as mentioned above, the  thermalization rate
may be smaller than
the expansion rate,
allowing for significant redshifting of the total energy density, leading to a
low reheat temperature, perhaps of the order of the inflaton mass or less.

The particular case of large thermalization rates compared to the expansion
rate, allows for
a quick estimate of the reheating temperature. In this case we can assume
that thermalization
occurs without red-shifting of the energy which is then conserved during the
thermalization time.
By taking the energy density at the
beginning of this stage to be $\alpha T^4$ with $\alpha$ the Stephan Boltzmann
accounting for the particle statistics one can then obtain an estimate of
the reheating
temperature, but the justification for this should ultimately arise from a
deeper understanding
of the time scales involved.

Our main point in all of this, though, is that  self-consistent
calculations taking into account the non-linearity of the dynamics
are necessary in order to extract the reheating temperature in a
reliable way. Our work clarifies what the relevant
particle production mechanisms are in these theories and  provide a
consistent and
implementable framework of calculation. We are now extending these
studies to FRW  cosmologies.

We emphasized in this study
that thermalization and particle production are fundamentally
different processes and in the
non-linear regime likely to occur on widely different time scales.

\subsection{\bf A possible solution to the Polonyi problem:}

 As mentioned before a problem of much  current interest in which this work
may have some implications is the ``post-modern Polonyi problem'' concerning
flat
directions for some of the moduli fields in string theories\cite{moduli}.
In these scenarios the Polonyi fields are very light (of order of a Tev)
with couplings to
the particle physics sector that is extremely small (order of $10^{-17}$).
A calculation of the reheating temperature
based on the ``old'' scenario of one particle decay give
$T_{reh} \approx 0.01 \mbox{ Mev }$, well below the nucleosynthesis scale.

This may not necessarily be true given our analysis. Since particle production
occurs not just during oscillations around the minimum now but throughout the
evolution of the the moduli field. Even for couplings
${\cal{O}}(m_{\phi}/M_{pl})$
since the initial amplitudes of these moduli field are ${\cal{O}}(M_{pl})$
then the
mechanism of particle production via induced amplification will be very
efficient and our
analysis leads to an estimate of the reheating temperature
 $T_{reh} \approx m_{\phi} \approx 1 \mbox{ Tev }$ that is larger than the
electroweak scale, alleviating the Polonyi problem. Thus this mechanism
may very
well provide a solution to the Polonyi problem reconciling moduli field
cosmology with
Standard Big Bang cosmology.

\acknowledgements

D. B. would like to thank F. Cooper, and R. Pisarski for illuminating
conversations.
D.B.  and D.-S. Lee thank the N.S.F. for support under grant awards:
PHY-9302534 and INT-9216755.
D.B., R.H. and D.-S. Lee would like to thank E. Mottola and A. Linde
for interesting
discussions. H. J. de V.  would like to thank I. Krichever , A. Linde
and N. S\'anchez for useful discussions.
D.S. thanks H.L. Yu and B.L. Hu for enlightening conversations.
H. J. de V. has been supported in part by a CNRS-NSF binational grant.
R. Holman was supported in part by DOE contract DE-FG02-91-ER40682.
M. D'A. would like to thank LPTHE (Paris VI-VII) for kind hospitality.

\appendix

\section{\bf A pedagogical exercise}
In this appendix we show explicitly how to implement the tadpole method for the
case of the inflaton coupled to a lighter scalar field via a simple trilinear
coupling. We will not worry about renormalization issues here, though they
were, of course, dealt with in the text.  The formalism for non-equilibrium
quantum field theory has already been described several times in the
literature\cite{noneq}. A path integral representation involves an integration
along a path in the complex time plane with forward, backward branches and if
the initial density matrix was that of an equilibrium system at an (initial)
temperature T also an imaginary time branch. For real time correlation
functions the imaginary time branch does not contribute and only determines the
boundary conditions on the Green's functions.

We use the tadpole method \cite{tadpole,reheating,elmfors} to study
the time evolution
of the expectation value of the inflaton.
\begin{equation}
\phi(\vec{x},t)\equiv
\frac{{\mbox{Tr\,}}[\Phi^+(\vec{x})\rho(t)]}{{\mbox{Tr\,}}\rho(0)}=
\frac{{\mbox{Tr\,}}[\Phi^-(\vec{x})\rho(t)]}{{\mbox{Tr\,}}\rho(0)}\;,
\nonumber
\end{equation}
where $\Phi^{\pm}$ are the fields defined on the forward and backward branches
respectively. We set
\begin{equation}
\Phi^{\pm}(\vec{x},t)=\phi(\vec{x},t)+\chi^{\pm}(\vec{x},t)\;,\nonumber
\end{equation}
where $\chi^{\pm}$ are field fluctuation operators defined along the respective
branches.

The effective evolution equation for the background field $\phi(\vec{x},t)$
follows from the condition.
\begin{equation}
<\chi^{\pm}(\vec{x},t)>=0\;. \label{tad}
\end{equation}
Treating the {\it linear} and non-linear terms in $\chi^{\pm}$ as interactions
and imposing the condition (\ref{tad}) consistently in a perturbative or loop
expansion, one obtains expressions of the form (here we quote the equation
obtained from $<\chi^{+}(\vec{x},t)>=0$ )
\begin{equation}
\int d\vec{x}'dt' \left[<\chi^{+}(\vec{x},t)
\chi^{+}(\vec{x}',t') >
{\cal{O}}^{++}(\vec{x}',t') +
<\chi^{+}(\vec{x},t) \chi^{-}(\vec{x}',t') > {\cal{O}}^{+-
}(\vec{x}',t')
\right] = 0
\nonumber
\end{equation}
and similarly for  $<\chi^{-}(\vec{x},t)>=0$. The ${\cal{O}}^{\pm \pm}$
are in general integro-differential operators acting on the background field.

Because the Green's functions $<\chi^{+}(\vec{x},t) \chi^{+}(\vec{x}',t') >$,
etc. are all independent one obtains the equations of motion in the form
\begin{equation}
{\cal{O}}^{++}(\vec{x}',t')=0\;, ~~ {\cal{O}}^{+-}(\vec{x}',t') =0\;, ~~
{\cal{O}}^{-+}(\vec{x}',t')=0\;, ~~ {\cal{O}}^{--}(\vec{x}',t') =0\;.
\nonumber
\end{equation}
It is a consequence of the properties of the non-equilibrium Green's function
(see below) and ultimately a consequence of unitarity that all the
integro-differential operators ${\cal{O}}$ are the same.

Consider the Lagrangian density
\begin{equation}
{\cal{L}}(\Phi,\sigma) = {\cal{L}}_0(\Phi)+{\cal{L}}_0 (\sigma)+
g(t)\; \Phi \; \sigma^2 
\nonumber\;,
\end{equation}
with the ${\cal{L}}_0$ being the free field Lagrangian density (with respective
mass terms) and have allowed the coupling to depend on time. The
non-equilibrium path integral requires ${\cal{L}}(\Phi^+,\sigma^+)-
{\cal{L}}(\Phi^-,\sigma^-)$.  In the tadpole method we write
$\Phi^{\pm}(\vec{x},t)= \chi^{\pm}(\vec{x},t)+\phi(t)$ and identify $\phi(t)$
as the (non-equilibrium) expectation value of the field $\Phi$ . This
identification then requires that $<\chi(\vec{x},t)>=0$ where the expectation
value is in the non-equilibrium density matrix with the path integral
representation along the contour in complex time. After this shift, the action
reads:
\begin{eqnarray}
L & = &  \int d^3x dt \left\{{\cal{L}}_0(\chi^+)+{\cal{L}}_0(\sigma^+)
+\chi^+\left[
-\ddot{\phi}-m^2_{\Phi}\phi \right]+g(t) \phi(t)(\sigma^+)^2 + g(t)
\chi^+(\sigma^+)^2
 \right. \nonumber \\
  &     & \left. -\left(\chi^+,\sigma^+
\rightarrow \chi^-, \sigma^-\right) \right\}\;. \label{extotlag}\nonumber
\end{eqnarray}
The linear term in $\chi^{\pm}$ is included as a perturbation. The first
contribution to the equation of motion is obtained from this linear term, from
the condition $<\chi^+(\vec{x},t)>=0$ one obtains to this order
\begin{equation}
\int dt' \left\{<\chi^+(t)\chi^+(t')>(i) \left[
-\ddot{\phi}-m^2_{\Phi}\phi \right] - <\chi^+(t)\chi^-(t')>(i) \left[
-\ddot{\phi}-m^2_{\Phi}\phi \right] \right\} = 0\;, 
\nonumber
\end{equation}
where we have suppressed the spatial arguments. Because the correlation
functions $<\chi^+(t)\chi^+(t')> ,~~ <\chi^+(t)\chi^-(t')>$ are independent,
one obtains the tree level equations of motion. It is straightforward to see
that the same is obtained by imposing $<\chi^-(\vec{x},t)>=0$. In an amplitude
expansion (an expansion in powers of $\phi(t)$) the one loop correction to the
equation of motion is obtained by expanding (the exponential of) $
g\phi(t)(\sigma^+)^2 + g\chi^+(\sigma^+)^2 - (+\rightarrow -)$ to first and
second order. The first order gives a tadpole contribution, the second order
needs one vertex with $\chi$, the other with $\phi$. One obtains
\begin{eqnarray}
&&\int dt' <\chi^+(t)\chi^+(t')>\left\{(i) \left(
-\ddot{\phi}-m^2_{\Phi}\phi \right)+(ig(t')) <(\sigma^+(t'))^2> +  \right.
\nonumber \\
&& \left.
\int dt'' (ig(t''))^2\left[<(\sigma^+(t'))^2(\sigma^+(t''))^2>-
<(\sigma^+(t'))^2(\sigma^-(t''))^2>
\right] \phi(t'') \right\}  - \nonumber \\
&& \int dt' <\chi^+(t)\chi^-(t')> \left\{(i) \left(
-\ddot{\phi}-m^2_{\Phi}\phi\right)+  (ig(t')) <(\sigma^-)^2(t')> - \right.
\nonumber \\
&& \left.
\int dt'' (-ig(t''))^2 \left[<(\sigma^-(t'))^2(\sigma^-(t''))^2>\phi(t'')
- <(\sigma^-(t'))^2(\sigma^+(t''))^2>\right]\phi(t'') \right\}
 = 0 \;. \label{secondord}\nonumber
\end{eqnarray}
The expectation values are computed using Wick's theorem and using the
free-field Green's functions of section II. The tadpole (time independent) is
absorbed in a shift of the expectation value. The coefficient of $<\chi^+
\chi^+>, ~~ <\chi^+ \chi^->$ must vanish independently because these Green's
functions are independent and must vanish at all times. From the Green's
functions (and more generally from the formal time contour integral) it is seen
that the equations obtained are identical. If the expectation value is
translational invariance, the spatial integrals set the momentum transfer to
zero. For example using the zero temperature Green's functions of section I,
one finds that the term that has the non-local (in time) correlation functions
(last term) becomes
\begin{eqnarray}
&& \int dt'' (ig(t''))^2\left[<(\sigma^+(t'))^2(\sigma^+(t''))^2>-
<(\sigma^+(t'))^2(\sigma^-(t''))^2>
\right] \phi(t'') =  \nonumber \\
&& \quad\quad 2 \int dt''(ig(t''))^2 \phi(t'') \Theta(t'-t'') \int
\frac{d^3k}{(2\pi)^3}(-2i)
\frac{\sin \left[2\omega_k(t'-t'')\right]}{4\omega_k^2}\;.
\label{exer1}\nonumber
\end{eqnarray}
For the resummed one-loop approximation the term $ g\phi(t)(\sigma^{\pm})^2 $
is absorbed in the mass term for the $\sigma$ field and only the $\chi\sigma^2$
terms are considered as perturbation. In this case the 1-loop contribution is
obtained from the tadpole term $<(\sigma^{\pm}(t))^2>$, which is now time
dependent and obtained from the non-equilibrium Green's functions constructed
with the mode functions for the time-dependent mass as in section IV. One now
finds the evolution equations
\begin{eqnarray}
&& \ddot{\phi}(t)+ m^2_{\Phi}\; \phi(t)+  g(t) <\sigma^2(\vec{x},t)>
=0\;,  \nonumber \\
&& <\sigma^2(\vec{x},t)> = -i\int \frac{d^3k}{(2\pi)^3}G^{++}_{k,\sigma}(t,t) =
\int \frac{d^3k}{(2\pi)^3} \frac{|V_k(t)|^2}{2w_k}
\coth \left[\frac{\beta w_k}{2}\right]\;,  \nonumber \\
&&\left[\frac{d^2}{dt^2}+\vec{k}^2+m^2(t)\right] V_k(t)=0\;, \nonumber   \\
&& V_k(0)=1\;, ~~~ \dot{V}_k(0)= -i w_k= -i \sqrt{\vec{k}^2+m^2(0)}\;,
\nonumber\\
&&m^2(t) = m_{\sigma}+g(t)\phi(t) \;.\nonumber
\end{eqnarray}

\newpage

{\bf Figure Captions:}

\underline{\bf Fig.(1.a) Scalars: unbroken symmetry}
$\eta(\tau)$ vs $\tau$ for the values of the parameters
$y=0;~~ \lambda /8\pi^2=0.2;~~g / \lambda=1; ~
 m_{\sigma}=0.2\,m_{\phi};~~ \eta(0)=1.0;~~\dot{\eta}(0)=0$.

\underline{\bf Fig.(1.b):} ${\cal{N}}_{\sigma}(\tau)$ vs. $\tau$ for the same
value of the parameters as figure (1.a).

\underline{\bf Fig.(1.c):}
${\cal{N}}_{q,\sigma}(\tau=120)$ vs. $q$ for the same
values as in figure (1.a).

\underline{\bf Fig.(2.a) Scalars: broken symmetry}
$\eta(\tau)$ vs $\tau$ for the values of the parameters
$y=0;~~ \lambda /8\pi^2=0.2;~~g / \lambda=0.05; ~
 m_{\sigma}=0.2\,|m_{\phi}|;~~ \eta(0)=0.6;~~\dot{\eta}(0)=0$.

\underline{\bf Fig.(2.b):} ${\cal{N}}_{\sigma}(\tau)$ vs. $\tau$ for the same
value of the parameters as figure (2.a).

\underline{\bf Fig.(2.c):} ${\cal{N}}_{q,\sigma}(\tau=200)$ vs. $q$
for the same
values as in figure (2.a).

\underline{\bf Fig.(3.a) Fermions: unbroken symmetry}
$\eta(\tau)$ vs $\tau$ for the values of the parameters
$g=0;~~y^2/\pi^2=0.5;~~ \lambda /6y^2=1;~~m_{\psi}=0; ~
\eta(0)=1.0;~~\dot{\eta}(0)=0$.

\underline{\bf Fig.(3.b):} ${\cal{N}}^f (\tau)$ vs. $\tau$ for the same
value of the parameters as figure (3.a).

\underline{\bf Fig.(3.c):} ${\cal{N}}^f_{q}(\tau=200)$ vs. $q$ for the same
values as in figure (3.a).

\end{document}